\documentclass[aps,pra,floatfix,twocolumn,superscriptaddress,showpacs,10pt]{revtex4-1}
\usepackage{amssymb,amsmath,color,mciteplus,graphicx,subfigure}
\usepackage{calc,epsfig,epstopdf,color,mciteplus,bm,mathrsfs}
\usepackage{times}
\usepackage{float}
\usepackage{lipsum}
\usepackage{color}
\usepackage{amsmath}
\usepackage[title]{appendix}

\usepackage[unicode=true,
bookmarks=true,bookmarksnumbered=false,bookmarksopen=false,
breaklinks=false,pdfborder={0 0 1},backref=false,colorlinks=true]
{hyperref}
\hypersetup{linkcolor=blue, urlcolor=blue, citecolor=blue,pdfstartview={FitH}, hyperfootnotes=false, unicode=true}

\newcommand\red[1]{{\color{red}#1}}
\begin{document}

\title{Leakage Suppression in Quantum Control via Static Parameter Offsets}

\author{Ting Lin} 
\affiliation{Key Laboratory of Atomic and Subatomic Structure and Quantum Control (Ministry of Education), Guangdong Basic Research Center of Excellence for Structure and Fundamental Interactions of Matter, and School of Physics, South China Normal University, Guangzhou 510006, China}

\author{Zi-Hao Qin} 
\affiliation{Key Laboratory of Atomic and Subatomic Structure and Quantum Control (Ministry of Education), Guangdong Basic Research Center of Excellence for Structure and Fundamental Interactions of Matter, and School of Physics, South China Normal University, Guangzhou 510006, China}

\author{Zheng-Yuan Xue} \email{zyxue83@163.com}
\affiliation{Key Laboratory of Atomic and Subatomic Structure and Quantum Control (Ministry of Education), Guangdong Basic Research Center of Excellence for Structure and Fundamental Interactions of Matter, and School of Physics, South China Normal University, Guangzhou 510006, China}
\affiliation{Guangdong Provincial Key Laboratory of Quantum Engineering and Quantum Materials, \\ Guangdong-Hong Kong Joint Laboratory of Quantum Matter, and Frontier Research Institute for Physics, \\  South China Normal University, Guangzhou 510006, China}

\author{Tao Chen} \email{chentamail@163.com}
\affiliation{Key Laboratory of Atomic and Subatomic Structure and Quantum Control (Ministry of Education), Guangdong Basic Research Center of Excellence for Structure and Fundamental Interactions of Matter, and School of Physics, South China Normal University, Guangzhou 510006, China}
\affiliation{Guangdong Provincial Key Laboratory of Quantum Engineering and Quantum Materials, \\ Guangdong-Hong Kong Joint Laboratory of Quantum Matter, and Frontier Research Institute for Physics, \\  South China Normal University, Guangzhou 510006, China}


\begin{abstract}
High-fidelity quantum operations require the system dynamics to be strictly confined to the computational subspace. In practice, however, control fields inevitably couple to leakage levels, giving rise to quantum state leakage that significantly reduces the fidelity of the operation. To address this challenge, we propose a general strategy for actively suppressing leakage errors by applying small, static offsets to tunable system parameters. This approach systematically mitigates leakage's detrimental impact on quantum control, without modifying the original control framework or incurring additional time overhead. By avoiding the need for extra suppression pulses or complex optimization procedures altogether, it offers a streamlined solution for leakage compensation while remaining fully compatible with subsequent optimal control techniques. Numerical validation conducted on superconducting quantum circuits demonstrates effective leakage suppression, enabling high-fidelity single-qubit gates, precise control of two-qubit interactions, and perfect state transfer in multi-level systems. Moreover, when integrated with optimal control techniques, our approach also allows for the cooperative suppression of both leakage errors and residual crosstalk. Therefore, this work provides a feasible technical pathway toward the low error thresholds required for fault-tolerant quantum computation.
\end{abstract}

\maketitle

\section{Introduction}

Quantum control \cite{QCQI}, as the core technology for achieving high-precision manipulation of quantum systems, directly determines the performance of quantum gates and the execution efficacy of quantum algorithms. In an ideal scenario, the dynamical evolution of a quantum system should be strictly confined to its computational subspace. However, in realistic physical implementations, unavoidable couplings between control fields and leakage levels lead to uncontrolled population of quantum states outside the computational space \cite{NISQ1, NISQ2}. Such leakage errors not only significantly reduce gate fidelity but can also induce and propagate correlated errors among qubits, thereby critically undermining the effectiveness of quantum error correction protocols \cite{QEC}. As the scale of quantum processors continues to grow, leakage error has become one of the key challenges that must be resolved to achieve fault-tolerant quantum computation. Consequently, the development of efficient leakage suppression techniques is of paramount importance for enhancing the overall performance of quantum computing systems.

To address the critical challenge of leakage errors in quantum control, a wide array of strategies have been developed, spanning theoretical principles to hardware-level implementations. A major category involves the optimization of control pulses \cite{PulseOP1, PulseOP2, PulseOP3, PulseOP4, PulseOP5, PulseOP6, PulseOP7, PulseOP8, PulseOP9, PulseOP10, PulseOP11, PulseOP12}, whose core idea is to actively suppress leakage transitions through carefully designed control fields while achieving target quantum gates. Prominent examples include the Gradient Ascent Pulse Engineering (GRAPE) algorithm \cite{PulseOP1}, which generates leakage-robust control pulses via numerical optimization, and analytical schemes such as the Derivative Removal by Adiabatic Gate (DRAG) \cite{PulseOP2, PulseOP3}, which counteracts dominant leakage channels by introducing specifically designed compensatory pulse components. Regarding control sequences, dynamical decoupling techniques \cite{DD1, DD2, DD3} utilize rapid pulse flipping to decouple systems from leakage energy levels, while composite pulses \cite{CP1, CP2, CP3, CP4} achieve self-compensation for leakage through a series of structured elementary pulses. Furthermore, dedicated leakage reduction units \cite{LRU1, LRU2, LRU3, LRU4} offer a hardware-efficient post-processing strategy that resets leakage states to the computational basis in real time, introducing minimal timing overhead. On the hardware front, hardware elements such as tunable couplers \cite{Coupler1, Coupler2, Coupler3, Coupler4, Coupler5} can be engineered to mitigate leakage by tailoring the energy level structure and interaction strengths of the system.

Here, we propose a general strategy for actively suppressing leakage errors by applying small, static offsets to tunable system parameters. This approach systematically mitigates leakage's detrimental impact on quantum control, without modifying the original control framework or incurring additional time overhead. By avoiding the need for extra suppression pulses or complex optimization procedures altogether, it offers a streamlined solution for leakage compensation while remaining fully compatible with subsequent optimal control techniques. To validate the effectiveness of our leakage suppression scheme, we conduct comprehensive testing on superconducting quantum circuits. The numerical results demonstrate that our approach enables high-fidelity single-qubit gate operations, precise control over two-qubit interactions, and perfect state transfer in multi-level systems. Moreover, when integrated with optimal control techniques, our approach facilitates the cooperative suppression of both leakage errors and residual crosstalk, thereby enhancing the overall performance of complex quantum circuits. Therefore, this work establishes a practical and efficient route to robust, high-fidelity quantum operations.

\section{THE GENERAL SCHEME}\label{GF}
Consider a quantum system with a Hilbert space $\mathbf{H}$ that can be partitioned into a computational subspace $\mathbf{H}_{\text{comp}}$ and a leakage subspace $\mathbf{H}_{\text{leak}}$. The computational subspace encompasses the logical states of the qubits. Ideally, when the system evolves under the target unitary operator $U_{\text{targ}}$, the quantum state should remain strictly confined within the computational subspace, achieving the desired target state with perfect accuracy. However, in realistic implementations, weak yet non-negligible couplings generally exist between $\mathbf{H}_{\text{comp}}$ and $\mathbf{H}_{\text{leak}}$. These couplings allow the population to escape into $\mathbf{H}_{\text{leak}}$, thereby reducing the gate fidelity.

For concreteness, we consider a system with $N$ basis states, where $\mathcal{K}$ of these states encode the logical states of qubits, forming the computational subspace. The remaining $N-\mathcal{K}$ basis states constitute the leakage subspace. The Hamiltonian of the system is expressed as \begin{equation}
\mathcal{H}(t) = \mathcal{H}_{\text{targ}}(t) + \mathcal{H}_{\text{leak}}(t),
\end{equation}
where $\mathcal{H}_{\text{targ}}(t)$ denotes the ideal Hamiltonian corresponding to the target evolution, including the effective dynamics within the computational subspace and the free terms within the quantum system. $\mathcal{H}_{\text{leak}}(t)$ describes the coupling between the computational and leakage subspaces, as well as interactions among the leakage levels themselves.

We defined a unitary time evolution operator as $U_{\text{all}}(t)=U_{\text{targ}}(t)U_{\text{err}}(t)$ \cite{PulseOP5}. 
Here, $U_{\text{targ}}(t)$ denotes the target evolution operator generated by the ideal Hamiltonian $\mathcal{H}_{\text{targ}}(t)$, defined as $U_{\text{targ}}(t)=\hat{\mathcal{T}}\exp\left[-i\int_0^T\mathcal{H}_{\text{targ}}(t)dt\right] $ where $\hat{\mathcal{T}}$ denotes the time-ordering operator and $T$ represents the overall evolution time. $U_{\text{err}}$ is generated by the form of the leakage Hamiltonian within the interaction picture, $ U_{\text{err}}(t)=\hat{\mathcal{T}}\exp\left[-i\int_0^T\mathcal{H}_{\text{I}}(t)dt\right]$, with $\mathcal{H}_{\text{I}}=U_{\text{targ}}^{\dagger}(t)\mathcal{H}_{\text{leak}}U_{\text{targ}}(t)$. We assume that the system undergoes an ideal leakage-free evolution throughout its dynamics. That is, at the final time $t = T$, the overall evolution operator $U_{\text{all}}(T)$ coincides with the target unitary operator $U_0$, completing the ideal quantum gate operation. To achieve this, the computational subspace must be decoupled from the leakage subspace. That is, the evolution generated by $\mathcal{H}_{\text{I}}(t)$ is equivalent to the identity operator within the computational subspace: 
\begin{align}
    \rho_\mathcal{K}U_{\text{err}}(t)\rho_{\mathcal{K}} = \mathbb{I}_\mathcal{K},
\end{align}
where $\rho_\mathcal{K}$ is the projection operator in the computational subspace and $\mathbb{I}_\mathcal{K}$ is the identity operator. 
Consequently, the overall evolution operator at the final time strictly corresponds to the following ideal unitary operation:
\begin{align}
    \rho_\mathcal{K}U_{\text{all}}(T)\rho_\mathcal{K}
    = \rho_\mathcal{K}U_{\text{targ}}(T)\rho_\mathcal{K}
    = U_0.
\end{align}


Observing the Hamiltonian reveals that the extent of leakage influence correlates with parameters associated with interactions between computational and leakage subspaces, as well as interactions among the leakage subspaces themselves. Importantly, the applied offsets for tunable parameters remain within a modest and experimentally accessible range, ensuring our approach does not introduce additional experimental complexity. Here, we aim to apply small and static offsets to the system's tunable parameters $\delta_{p_i}=\{\delta_{p_1},~\delta_{p_2},~\delta_{p_3}\dots\}$ to counteract the dynamical effects of the system's passive leakage terms. Consequently, the Hamiltonian can be rewritten as
\begin{subequations} 
\begin{align}
     \mathcal{H}(t)&\rightarrow\mathcal{H}(t,\delta_{p_i}), \\
     \mathcal{H}_{\text{targ}}(t)+\mathcal{H}_{\text{leak}}(t)&\rightarrow\mathcal{H}_{\text{targ}}(t,\delta_{p_i})+\mathcal{H}_{\text{leak}}(t,\delta_{p_i}).\label{Hdelta}
\end{align}
\end{subequations}


To determine the form of the offsets to the Hamiltonian parameters for suppressing leakage errors, we rotate the system into the unitary transformation associated with $\delta_{p_i}$ with transfer matrix $A(\delta_{p_i})$, 
\begin{equation}
     \mathcal{H}_A(t)=A^\dagger(\delta_{p_i}) \mathcal{H}(t)A(\delta_{p_i})+i\dot{A}(\delta_{p_i})A(\delta_{p_i}).
\end{equation}
Designing an appropriate $A(\delta_{p_i})$ is crucial, as it ensures that the evolution operator generated by $\mathcal{H}_A(t)$ implements the ideal gate operation at the final time, i.e.,
\begin{equation}
    U_A(T,0)=\hat{\mathcal{T}}e^{-i\int_0^T \mathcal{H}_A(t)dt}=U_0.
\end{equation}

We divide the evolution time into $k$ segments, where $k$ is an integer, i.e., $T = k\tau$. Accordingly, $U_A(T,0)$ can be expressed as
\begin{equation}
    U_A(T,0) = \prod_{n=1}^{k} U_A[n\tau, (n-1)\tau].
\end{equation}
The target evolution can be achieved as long as the contribution of leakage errors in each time segment is zero. For the $n$-th time segment, the evolution operator can be written as $U_A[n\tau, (n-1)\tau] = \hat{\mathcal{T}} \exp\big[-i \int_{(n-1)\tau}^{n\tau} \mathcal{H}_A(t)\, dt \big].$ Using the Magnus expansion \cite{Magnus1, Magnus2}, the time-ordered exponential can be expressed as a simple exponential, so that the evolution operator for the $n$-th segment becomes
\begin{equation}
    U_A[n\tau, (n-1)\tau] = \exp\Big[-i\tau \big(\bar{\mathcal{H}}^{(0)} + \bar{\mathcal{H}}^{(1)} + \dots \big) \Big].
\end{equation}

Based on the first-order approximation, the above evolution operator for the $n$-th time segment can be expressed as $U_A[n\tau, (n-1)\tau] = \exp[-i\tau \bar{\mathcal{H}}^{(0)}]$, i.e.,
\begin{align}
    \bar{ \mathcal{H}}^{(0)}=&\frac{1}{\tau}\int^{n\tau}_{(n-1)\tau} \mathcal{H}_A(t)dt\notag\\
    =&\frac{1}{\tau}\bigg[\int_{(n-1)\tau}^{n\tau} \mathcal{H}_{A0}(t)dt\notag\\
    &+\int_{(n-1)\tau}^{n\tau}A^\dagger(\delta_i) \mathcal{H}_{leak}(t)A(\delta_i)dt\bigg], 
\end{align}
with
\begin{equation}
    \mathcal{H}_{A0}(t) = A^\dagger(\delta_{p_i}) \mathcal{H}_0 A(\delta_{p_i}) + i \dot{A}^\dagger(\delta_{p_i}) A(\delta_{p_i})
\end{equation}
being the leakage-free Hamiltonian in the new picture, which can generate the ideal gate $U_0$. To eliminate the effect of leakage errors, the following two boundary conditions must be satisfied:
\begin{subequations} 
\begin{align}
    A&(t, \delta_{p_i}) = A(t+\tau, \delta_{p_i}),\label{cond1}\\
    \int_0^{\tau} A^\dagger&(\delta_{p_i}) \mathcal{H}_{\rm leak}(t) A(\delta_{p_i})\, dt = 0.\label{cond2}
\end{align}
\end{subequations}
Constructing $A(\delta_{p_i})$ to satisfy these boundary conditions allows for effective suppression of leakage errors. To evaluate the effectiveness of the optimization, we use the gate fidelity calculation formula, defined as $F^\text{G}=\text{Tr}(U_fU_0^\dagger)/\text{Tr}(U_0U_0^\dagger)$, where $U_0$ represents the target gate operation, while $U_f$ denotes the evolution operator generated by the corrected Hamiltonian $\mathcal{H}(t, \delta_{p_i})$. 
We define the gate infidelity as $\text{InF}^\text{G}(\delta_{p_i}) = 1 - F^\text{G}$, which is influenced by the applied offsets and may include both positive and negative contributions. The optimization objective is to adjust the small, static offsets $\delta_{p_i} = \{\delta_{p_1}, \delta_{p_2}, \dots\}$ such that their impact on infidelity is minimized. 

We have structured the first few steps of small, static offsets into a general algorithm. Initially, identify the target gate and the corresponding system model. Given the target gate $U_0$, derive the Hamiltonian of the system model associated with this gate. On this basis, specify the target Hamiltonian $\mathcal{H}_{\text{targ}}$ for the implementation of the gate, as well as the leakage Hamiltonian or its primary leakage terms $\mathcal{H}_\text{leak}$. Secondly, construct the transformation matrix that satisfies two constraint conditions. Based on the system Hamiltonian for implementing the target gate, determine the transformation matrix $A(\delta_{q})$ (with $q=1,2,3,...$) that fulfills the constraints defined by Eq.~(\ref{cond1}) and Eq.~(\ref{cond2}). Here, $\delta_q$ represents the tunable transformation parameter. This step effectively converts the boundary conditions presented into constraints on $\delta_q$. Thirdly, establish the correspondence between the transformation parameters and the tunable parameters. Utilizing the transformation matrix $A(\delta_{q})$, which is modulated based on the tunable transformation parameter $\delta_q$, rotate the system Hamiltonian to obtain the Hamiltonian corresponding to the leakage-free system. Furthermore, clarify the relationship between each transformation parameter $\delta_q$ in $A(\delta_q)$ and the tunable parameters $\delta_{p_i}$ in the Hamiltonian, thereby deriving the expression provided in Eq.~(\ref{Hdelta}). This step maps the transformation parameters $\delta_q$ to the static experimentally tunable parameter offsets $\delta_{q_i}$. Finally, perform numerical validation and effectiveness assessment. The objective is to conduct numerical validation using gate fidelity as the target function, aiming to minimizing the impact of leakage errors on gate fidelity. Compare the evolution results before and after the introduction of parameter offsets to confirm that the selected tunable parameters effectively suppress leakage errors.

It is worth noting that our static parameter offset strategy does not rely on the specific energy-level structure or coupling form of a given system. As long as the leakage coupling terms are associated with the tunable control parameters in the target ideal Hamiltonian, leakage errors can be effectively suppressed by introducing small, static offsets to tunable system parameters. This indicates that our scheme can, in principle, be extended to different quantum platforms with similar control capabilities, such as superconducting quantum circuits, trapped ions, semiconductor quantum dots, and neutral atom systems, among others. These platforms already possess mature parameter control techniques, thus facilitating experimental implementation. The scheme is not only applicable to single-qubit quantum manipulation but can also be extended to two-qubit quantum control, and perfect state transfer in multi-level systems. Furthermore, since our approach does not require modifying the time-dependent shape of the original control pulses, it inherently retains the potential for integration with various pulse-shaping-based optimal control techniques. This enables cooperative suppression of other key error sources on each platform while simultaneously mitigating leakage. In the following, we will take superconducting quantum circuits as an example to demonstrate the applicability and effectiveness of this approach.


\section{EXAMPLES OF UNIVERSAL QUANTUM CONTROL} 

In this section, we demonstrate the effective suppression of leakage errors in superconducting quantum systems by applying small, static offsets to the tunable system parameters. This approach enables high-fidelity single-qubit operations and precise two-qubit control.

\subsection{Single-qubit Control}
\begin{figure}[tp]
    \centering
    \includegraphics[width=1.0\linewidth]{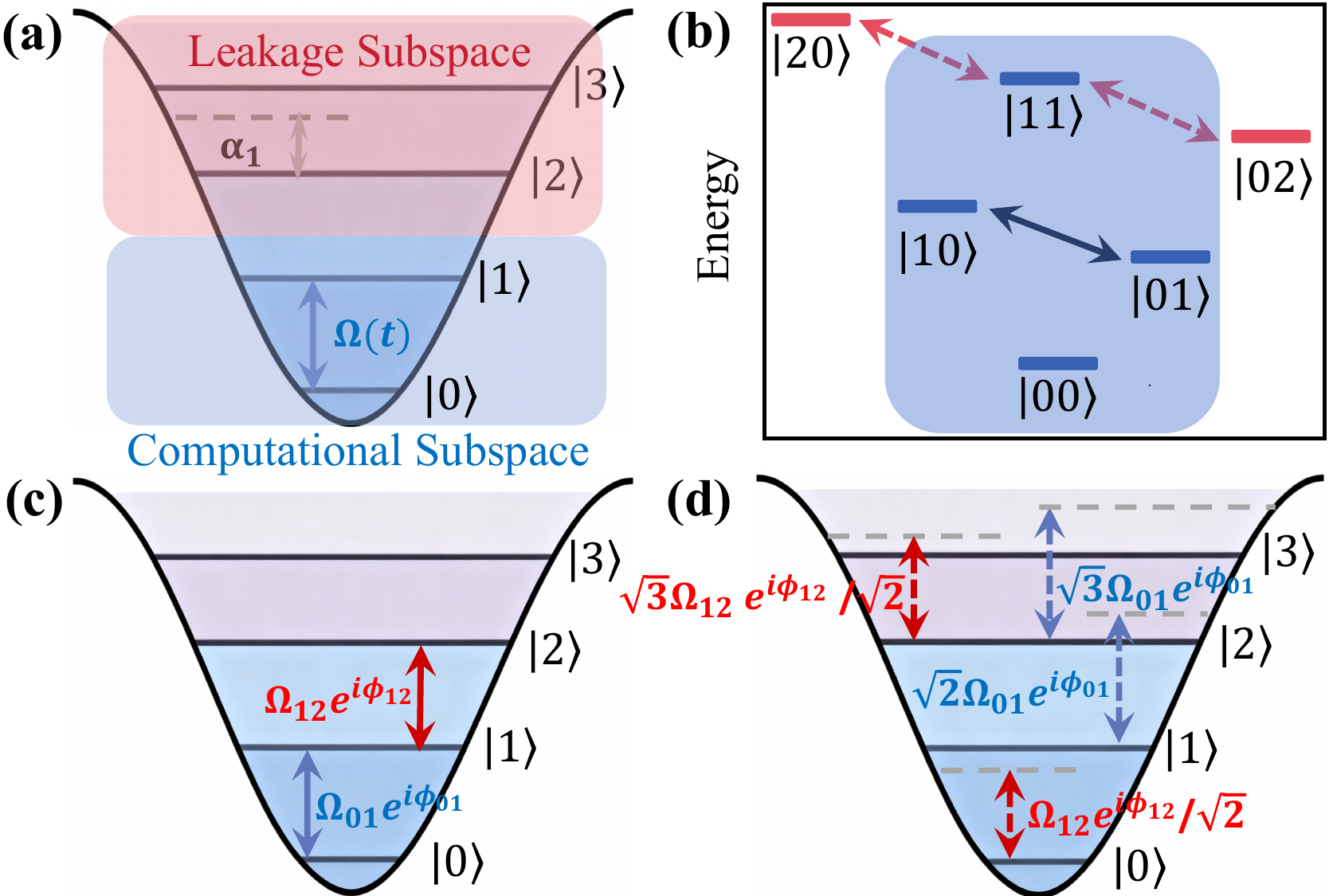}
    \caption{(a) The nonequidistant energy-spectrum diagram for a driven superconducting transmon-type qubit, in which $\{|0\rangle,~|1\rangle\}$ and $\{|2\rangle,~|3\rangle\}$ are taken as the computational subspace (blue box) and leakage subspace (red box), respectively. The dotted gray line indicates the positions the energy level would be at if the system was a harmonic oscillator. (b) Energy level diagram for two capacitively coupled transmon qubits, which can be used to implement the iSWAP gates. The dashed line indicates leakage errors between energy levels. (c) Two independent pulses are applied between energy levels $\{|0\rangle, |1\rangle\}$ (red) and $\{|1\rangle, |2\rangle\}$ (blue), thereby driving the target energy level transitions. (d) The dashed lines indicate leakage errors between energy levels outside the target energy level, with each color representing leakage errors caused by different pulse drivers.}
    \label{F1}
\end{figure}

\begin{figure*}
    \centering
    \includegraphics[width=0.9\linewidth]{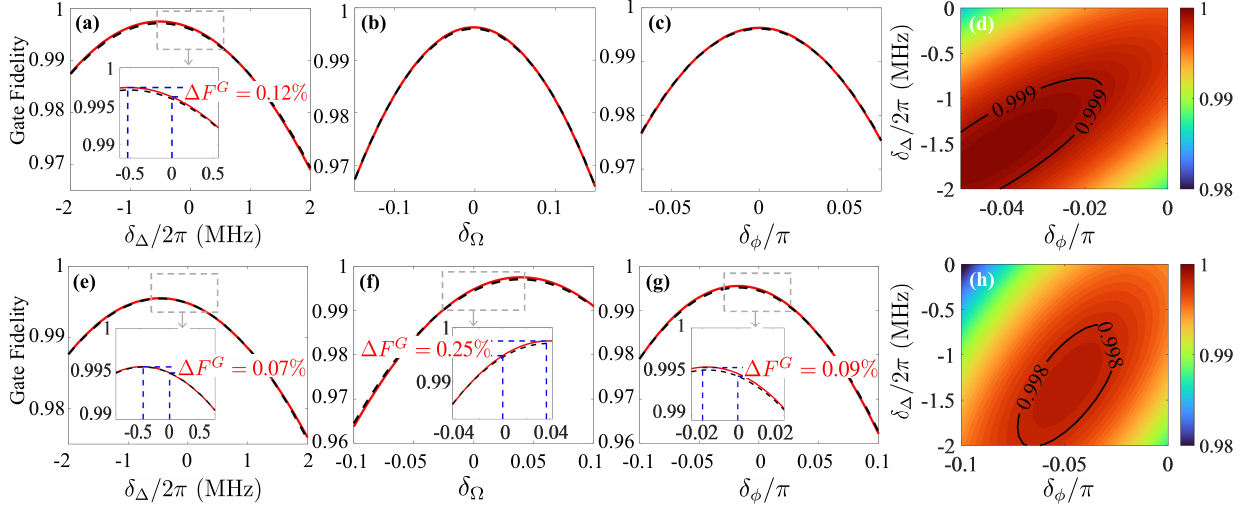}
    \caption{The approximate solutions from calculations (dashed lines) and exact solutions from numerical simulations (solid lines) for gate fidelity as functions of small, static offsets of coupling strength $\delta_\Omega$, phase $\delta_\phi$ and detuning $\delta_\Delta$, respectively, for NOT (above column) and Hadamard (below column) gates. $\Delta F^G$ denotes the enhancement in fidelity achieved by introducing offsets, as compared to without offsets.}
    \label{single_offset}
\end{figure*}

We consider a transmon-type superconducting single-qubit system \cite{Geo-SQ1, Geo-SQ2,Geo-SQ3,Geo-SQ4,Geo-SQ5,Geo-SQ6,Geo-SQ7}, where the logical qubit is encoded by the two lowest energy levels, denoted as $|0\rangle$ and $|1\rangle$. Ideally, quantum state evolution should remain confined to these two logical energy levels. However, non-ideal driving or coupling between energy levels can cause the quantum state to transition to state $|2\rangle$ or higher, resulting in leakage errors, as illustrated in Fig.~\ref{F1}(a). Therefore, suppressing these leakage transitions is essential for improving quantum control performance. When a driving pulse $\Omega(t)$ is applied to the superconducting qubit, the system's Hamiltonian can be expressed in the rotating reference frame at the driving frequency, using the rotating wave approximation, as
\begin{align}
    \mathcal{H}^1(t) = \mathcal{H}_{\text{targ}}^1(t) + \mathcal{H}_{\text{leak}}^1(t),
\end{align}
where, $\mathcal{H}_{\text{targ}}^1(t)$ governs the dynamics within the ideal logical subspace, corresponding to a standard Rabi oscillation process:
\begin{align}
    \mathcal{H}_{\text{targ}}^1(t) 
    =&\frac{\Delta}{2}\!\left(-|0\rangle\langle 0| + |1\rangle\langle 1|\right)
     + \left(\frac{3}{2}\Delta + \alpha\right)|2\rangle\langle 2| \notag \\
     &+ \big[\Omega(t)\,|0\rangle\langle 1| e^{i\phi(t)} + \text{H.c.}\big],
\end{align}
with $\Delta$ representing the detuning between the qubit and the pulse frequency, $\alpha$ denoting the anharmonicity, and $\phi$ indicating the pulse phase. Without loss of generality, we set $\Delta = 0$ and $\phi = 0$. On the other hand, the leakage Hamiltonian $\mathcal{H}_{\text{leak}}^1(t)$ describes the coupling between the computational subspace and the leakage subspace, specifically the non-ideal interaction between the $|1\rangle$ and $|2\rangle$ energy levels, and is written as
\begin{align}
    \mathcal{H}_{\text{leak}}^1(t)
    = \lambda\,\Omega(t)\,|1\rangle\langle 2| e^{i\phi(t)} + \text{H.c.},
\end{align}
where $\lambda$ characterizes the strength of the transition $|1\rangle \leftrightarrow |2\rangle$ relative to the transition $|0\rangle \leftrightarrow |1\rangle$, and can be set to $\sqrt{2}$. This term's presence induces transitions of the quantum state from the computational subspace to the leakage subspace, significantly reducing quantum gate fidelity. Throughout this work, we employ the pulse form $\Omega(t) = \Omega_m \sin(\pi t/T)$, where $\Omega_m$ denotes the peak amplitude of the pulse, and $T$ represents the pulse duration.

Next, we aim to identify an appropriate transformation operator to effectively suppress leakage errors. In particular, in a superconducting single-qubit system, we define the offset parameters as $\delta_{p_i} = \{\delta_{p_1}, \delta_{p_2}, \dots\} = \{\delta_x, \delta_y, \delta_z\}$. The transformation operator $A_1(\delta_{q,q=x,y,z})$ can be defined as
\begin{subequations}
    \begin{align}
    A_1(X_1) = &\exp\big[i X_1(\delta_{q}, q=x,y,z)\big],\label{a1}\\
    X_1(\delta_q) = &\frac{\delta_z}{2} \big(-|0\rangle\langle0| + |1\rangle\langle1| + 3|2\rangle\langle2|\big) \notag \\
    &+ \big[(\delta_x + i \delta_y)(|0\rangle\langle1| + \lambda |1\rangle\langle2|) + \text{H.c.}\big].
\end{align}
\end{subequations}
It is evident that $A_1(\delta_q)$ satisfies Eq. \eqref{cond1}, and the parameters $\delta_x$, $\delta_y$, and $\delta_z$ are used to satisfy Eq. \eqref{cond2}, with all three parameters being constants. Subsequently, the system Hamiltonian is rotated to a unitary transformation related to $\delta_{p_i}$, with the matrix of the unitary transformation given by Eq. \eqref{a1} as
\begin{align}
     \mathcal{H}_\text{I}^1(t)=&A_1(\delta_q) \mathcal{H}_1(t)A_1^{\dagger}(\delta_q)+i\dot{A_1}(\delta_q)A_1^{\dagger}(\delta_q)\notag\\
    = &\frac{\Omega\delta_y}{2}[-2|0\rangle\langle0|-2|1\rangle\langle1|+(2+\sqrt{2})|2\rangle\langle2|]+\alpha|2\rangle\langle2|\notag\\
    &+\frac{\sqrt{1+\delta_z^2}\Omega}{2} (|0\rangle\langle1|e^{-i\theta}+\sqrt{2}|1\rangle\langle2|e^{-i\theta}+\text{H.c.})\notag\\
    &+\frac{\sqrt{2}\alpha\Omega}{2}[(i\delta_x-\delta_y)|1\rangle\langle2|+\text{H.c.}], \label{h1}
\end{align}
where $\theta = \arccos(1/\sqrt{1 + \delta_z^2})$.  

For the Hamiltonian above, by appropriately adjusting the parameters $\delta_x$, $\delta_y$, and $\delta_z$, equivalently renders the coupling term $|1\rangle\langle2|$ in the transformed Hamiltonian can be made significantly smaller than the $|2\rangle\langle2|$ term, thereby effectively suppressing leakage errors. It is important to note that in superconducting single-qubit systems, small and static parameter offsets correspond to $\delta_\Omega$, $\delta_\Delta$, and $\delta_\phi$, which arise from the coupling strength, detuning, and phase in the system Hamiltonian. Therefore, applying these offsets can mitigate the impact of leakage errors. Consequently, in a superconducting single-qubit system, the offset parameters can be re-defined as $\delta_{p_i} = \{\delta_{p_1}, \delta_{p_2}, \delta_{p_3}\} = \{\delta_\Omega, \delta_\Delta, \delta_\phi\}$. For the offsets in coupling strength, it is expressed as $\Omega_m \rightarrow (1+\delta_\Omega)\Omega_m$, where $\delta_\Omega = \sqrt{1+\delta_z^2} - 1$; for the phase offsets, it is expressed as $\phi \rightarrow \phi + \delta_\phi$, where $\delta_\phi = \arccos(1/\sqrt{1+\delta_z^2})$; for the detuning offsets, $\Delta \rightarrow \Delta + \delta_\Delta$.

Here we use the typical examples of the NOT gate and the Hadamard gate for illustration. To determine the relationship between offset parameters and gate fidelity, we employ the formula $F^\text{G}=\text{Tr}(U_fU_0^\dagger)/\text{Tr}(U_0U_0^\dagger)$. Based on relevant studies \cite{parem1,parem2,parem3}, we choose conservative qubit parameters: a pulse peak value of $\Omega_m=2\pi\times30~\text{MHz}$ and a qubit anharmonicity of $\alpha=2\pi\times220~\text{MHz}$. The results of our calculations are as follows:
\begin{subequations}
    \begin{align}
       &F^G_{N}(\delta_\Omega) \approx -1.306\delta_\Omega^2 - 0.0045\delta_\Omega + 0.996,\\
       &F^G_{N}(\delta_\Delta) \approx -0.0044\bigg(\frac{\delta_\Delta}{2\pi}\bigg)^2 - 0.0045\frac{\delta_\Delta}{2\pi} + 0.996,\\
       &F^G_{N}(\delta_\phi) \approx -4.2966\bigg(\frac{\delta_\phi}{\pi}\bigg)^2 + 0.994;\\
       &F^G_{H}(\delta_\Omega) \approx -1.683\delta_\Omega^2 + 0.1317\delta_\Omega + 0.9945,\\
       &F^G_{H}(\delta_\Delta) \approx -0.0033\bigg(\frac{\delta_\Delta}{2\pi}\bigg)^2 - 0.003\frac{\delta_\Delta}{2\pi} + 0.9948,\\
       &F^G_{H}(\delta_\phi) \approx -2.372\bigg(\frac{\delta_\phi}{\pi}\bigg)^2 - 0.0824\frac{\delta_\phi}{\pi} + 0.9945.
    \end{align}
\end{subequations}

Figures~\ref{single_offset}(a)-\ref{single_offset}(c) and Figures~\ref{single_offset}(e)-\ref{single_offset}(g) illustrate the approximate solutions from calculations (dashed lines) and exact solutions from numerical simulations (solid lines) for gate fidelity as a function of offsets in coupling strength $\delta_\Omega$, detuning $\delta_\Delta$, and phase $\delta_\phi$ in the NOT and Hadamard gates. The calculated results align closely with the numerical simulations, demonstrating consistency. These findings reveal that each of the three parameters significantly influence gate fidelity, with both positive and negative effects. Fidelity improves when parameters are offset towards specific value ranges, resulting in a positive impact; conversely, offsetting parameters towards other ranges can negatively affect fidelity. Thus, it is essential to strategically optimize small and static offsets to enhance fidelity.

Notice that, Eq. \eqref{h1} reveals interdependencies among different offset parameters, prompting us to simultaneously adjust two or more parameters for optimization. Obviously, the simulation results in Fig.~\ref{single_offset}(d) and Fig.~\ref{single_offset}(h) effectively demonstrate that by simultaneously offsetting the detuning $\delta_\Delta$ and phase parameters $\delta_\phi$, further suppression of leakage errors can be achieved. Next, we simultaneously consider the adjustment of three parameters. For the NOT gate, setting the coupling strength offset to $\delta_\Omega=0.003$, the detuning offset to $\delta_\Delta=-2\pi\times1.55~\text{MHz}$, and the phase offset to $\delta_\phi=-0.040\pi$ resulted in an increase in gate fidelity from $99.63\%$ to $99.99\%$. For the Hadamard gate, with parameter offsets to $\delta_\Omega=-0.052,~\delta_\Delta=-2\pi\times2.95~\text{MHz}$, and $\delta_\phi=-0.114\pi$, the gate fidelity increased from $99.48\%$ to $99.95\%$. Importantly, the applied offsets are kept within a modest and experimentally accessible range, ensuring that our approach does not add experimental complexity. As we show in Appendix \ref{AppA}, the scheme still maintains a significant leakage suppression effect even at lower experimental precision.

\begin{figure}
    \centering
    \includegraphics[width=1.0\linewidth]{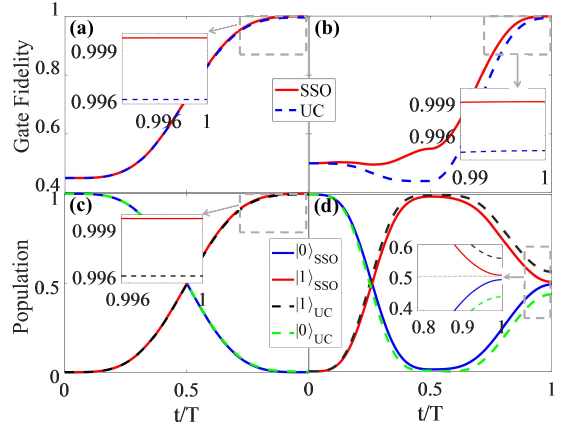}
    \caption{Comparison of the gate fidelity obtained with the optimized parameters under the small, static offsets (SSO) scheme and those uncorrected (UC). Figures (a) and (b) correspond to the NOT and Hadamard gates considering decoherence, respectively. The state population for NOT gate (c) and Hadamard gate (d) with the optimized parameters under the small, static offsets (SSO) scheme and those uncorrected (UC) while considering decoherence effect. The initial state of both gates is $|0\rangle$. }
    \label{single_fidelity}
\end{figure}

To thoroughly evaluate the impact of practical physical implementation, we must consider the effects of decoherence, as quantum systems inevitably coupled with their environment. Consequently, in subsequent numerical simulations, we account for the effects of decoherence, leakage terms, and small and static offsets on gate fidelity, thereby assessing the overall effectiveness of our scheme. For this purpose, we utilize the quantum master equation \cite{PulseOP2}:
\begin{align}
    \dot{\rho}_{q1}=&-i[\mathcal{H}_\text{I}^1(t,\delta_\Omega,\delta_\Delta,\delta_\phi),\rho_{q1}]+\frac{1}{2}\sum_{u=1,\varphi}\frac{\kappa_u}{2}\mathscr{L}(X_u^{q1}),
\end{align}
in which $\rho_{q1}$ is the density operator of the superconducting single-qubit system, and $\mathscr{L}(\mathcal{A}) = \mathcal{A}\rho\mathcal{A}^\dagger - \frac{1}{2}(\mathcal{A}^\dagger\mathcal{A}\rho + \rho\mathcal{A}^\dagger\mathcal{A})$ is the Lindblad operator associated with operator $\mathcal{A}$. Here, $X_1 = \sum_{j=0}^{+\infty}\sqrt{j+1}|j\rangle\langle j+1|$ and $X_\varphi = \sum_{j=0}^{+\infty}j|j\rangle\langle j|$. $\kappa_1$ and $\kappa_\varphi$ denote the decay rate and dephasing rate, respectively. Solving the master equation yields the final density matrix $\rho_{\text{q}1}$ of the single qubit, enabling a comprehensive evaluation of gate fidelity for the active leakage error suppression scheme. We define the gate fidelity of a single qubit as $F_1^G = \frac{1}{2\pi}\int_{0}^{2\pi}\langle\varPhi_{\text{q}1}|\rho_{\text{q}1}|\varPhi_{\text{q}1}\rangle d\theta_1$ \cite{singlefg1,singlefg2}, performing numerical integration over 1001 input states, with $\theta_1$ uniformly distributed in $[0,2\pi]$. Here, $|\varPhi_{\text{q}1}\rangle = U_0|\varPhi_1\rangle$ is the ideal final state of the general initial state of a single logical qubit $|\varPhi_1\rangle = \cos\theta_1|0\rangle + \sin\theta_1|1\rangle$. The NOT gate and Hadamard gate produce ideal final states $|\varPhi_{fN}\rangle = \cos\theta_1|1\rangle + \sin\theta_1|0\rangle$ and $|\varPhi_{fH}\rangle = (1/\sqrt{2})[(\cos\theta_1 + \sin\theta_1)|0\rangle + (\cos\theta_1 - \sin\theta_1)|1\rangle]$. Based on the latest experimental technology, we set conservative parameter regions \cite{parem1,parem2,parem3}, where the decay rate and dephasing rate of transmon qubits are $\kappa_1 = \kappa_\varphi = 2\pi\times2~\text{kHz}$. Figure~\ref{single_fidelity}(a) demonstrates that, considering decoherence, the fidelity of the NOT gate increases from $99.61\%$ to $99.98\%$ when the small and static offsets to system parameters are optimally selected. Similarly, Figure~\ref{single_fidelity}(b) demonstrates that, the fidelity of the Hadamard gate increases from $99.46\%$ to $99.92\%$. Furthermore, our scheme demonstrates robustness against calibration errors, as clearly illustrated in Appendix \ref{AppA}, which indicates that the scheme can maintain high fidelity across a broad range of error values. Additionally, the gate fidelity achieved by our scheme is comparable to that of DRAG, as detailed in Appendix \ref{AppB}.

Furthermore, we assume the qubit is initially in the state $|\psi_1\rangle=|0\rangle$, whereby the ideal NOT gate and Hadamard gate produce the final states $|\psi_{fN}\rangle=|1\rangle$ and $|\psi_{fH}\rangle=(|0\rangle+|1\rangle)/\sqrt{2}$, respectively. As shown in Fig.~\ref{single_fidelity}(c)-(d), the suppression of leakage errors through small, static offsets of tunable system parameters can also be assessed by examining the corresponding state populations.



\subsection{Two-qubit Control}

Single- and two-qubit gates constitute the fundamental operational units of universal quantum computation. 
In this section, we will continue to apply small, static offsets to system parameters to effectively suppress leakage errors in two-qubit gates. Non-trivial two-qubit gates can be implemented on two capacitively coupled transmon qubits, denoted as $T_1$ and $T_2$, with the corresponding Hamiltonian expressed in the following form: 
\begin{align}
    \mathcal{H}_c=&\sum_{k=1,2}\sum^2_{j=1}[j\omega_k-j(j-1)\alpha_k/2]\chi_j^k\notag\\
    &+\bigg[g_{12}\prod_{k=1,2}(\sum_{j=1}^2\lambda_j\sigma_j^k)+\text{H.c.}\bigg].
\end{align}
Here, $\chi_j^k = |j\rangle_k\langle j|$ denotes the projection operator onto the $j$-th level of qubit $k$, where the subscript $k$ is used to distinguish different qubits. The parameter $g_{12}$ signifies the coupling strength between qubits. The operator $\sigma_j^k=|j-1\rangle_k\langle j|$ serves as the standard lowering operator for qubits. The transition frequency is expressed as $j\omega-j(j-1)\alpha/2$, where $\alpha$ represents the intrinsic anharmonicity of the transmon qubits. The constant $\lambda_j$, which quantifies the strength of the transition $|j\rangle\leftrightarrow|j-1\rangle$, is determined by the dipole transition element. For generality, we define $\lambda_1=1$ and $\lambda_2=\sqrt{2}$.

The computational subspace of a two-qubit system consists of the states $|00\rangle,\,|01\rangle,\,|10\rangle,\,|11\rangle$. Ideally, the quantum state remains entirely within this subspace. However, similar to the single-qubit scenario, unavoidable coupling between different levels in real systems can cause the quantum state to transition to other levels, such as $|02\rangle$ and $|20\rangle$. These levels collectively form the leakage subspace, thereby introducing significant leakage errors, as illustrated in Fig.~\ref{F1}(b) This section aims to suppress leakage errors by actively adjusting parameters in the Hamiltonian. Typically, the coupling strength and frequency difference between adjacent transmon qubits are fixed and non-adjustable. To achieve more tunable parameters, we employ a parametrically tunable coupling by applying an AC drive to the transmon qubit $T_1$. Experimentally, this is achieved by biasing the qubit with an AC magnetic flux, allowing the transition frequency of the qubit to be periodically modulated \cite{omega1,omega2,omega3,omega4} as $ \omega_1(t)=\omega_1+\varepsilon_1\sin(\nu_1t+\phi_1)$. In the interaction picture, the Hamiltonian is expressed as follows:
\begin{align}
\mathcal{H}^2(t)=&g_{12}\{|01\rangle\langle10|e^{i\Delta_1t}e^{i\beta_1\cos(\nu_1t+\phi_1)}\notag\\
&+\sqrt{2}|02\rangle\langle11|e^{i(\Delta_1-\alpha_2)t}e^{i\beta_1\cos(\nu_1t+\phi_1)}\notag\\
&+\sqrt{2}|11\rangle\langle20|e^{i(\Delta_1+\alpha_1)t}e^{i\beta_1\cos(\nu_1t+\phi_1)}+\text{H.c.}\}
\end{align}
where $\Delta_1=\omega_2-\omega_1$, $\beta_1=\varepsilon_1/\nu_1$, and $|mn\rangle=|m\rangle_1\otimes|n\rangle_2$. Utilizing the Jacobi-Anger identity, and considering $J_m(\beta_1)$ as the Bessel function of the first kind, we set the parameter $\Delta_1-\nu_1=-\Delta_t$, with $|\Delta_t|\ll\{\nu_1, \Delta_1\}$. The system Hamiltonian is then expressed as: 
\begin{align}
    \mathcal{H}^2(t)=\mathcal{H}^2_{\text{targ}}(t)+\mathcal{H}^2_{\text{leak}}(t),
\end{align}
where $\mathcal{H}^2_\text{targ}(t)$ characterizes the dynamical evolution within the ideal computational subspace, representing an effective Rabi oscillation process. Its explicit form is given by: 
\begin{align}
\mathcal{H}_{\text{trag}}^2(t)=J_1(\beta_1)g_{12}\big[|01\rangle\langle10|e^{-i\Delta_tt}e^{-i(\phi_1-\frac{\pi}{2})}+\text{H.c.}\big]. 
\end{align}
By setting $\Delta_t=0$, $\phi_1=3\pi/2$, and $J_1(\beta_1)g_{12}\tau=\pi/2$, with $\tau$ being the pulse duration, a two-qubit $i$SWAP gate can be implemented. In contrast, $\mathcal{H}_{\text{leak}}^2(t)$ represents the interaction between the computational and leakage subspaces. The terms that significantly influence leakage are:
\begin{align}
     \mathcal{H}^2_\text{leak}(t)=&\sqrt{2}J_1(\beta_1)g_{12}\big[|02\rangle\langle11|e^{-i(\Delta_t+\alpha_2)t}e^{-i(\phi_1-\frac{\pi}{2})}\notag\\
    &+|11\rangle\langle20|e^{-i(\Delta_t-\alpha_1)t}e^{-i(\phi_1-\frac{\pi}{2})}+\text{H.c.}\big].
\end{align}
In the ideal $i$SWAP gate, the state $|11\rangle$ is affected solely by the identity operator. However, in practical physical systems, transitions to higher energy levels, such as $|02\rangle$ and $|20\rangle$, occur and serve as primary leakage sources that require attention.

We proceed by applying a unitary transformation to the leakage Hamiltonian,  with the matrix form given by $V=\exp\left\{i\left[\alpha_2|02\rangle\langle02|-\Delta_{e}|11\rangle\langle11|-(2\Delta_{e}-\alpha_1)|20\rangle\langle20|\right]t\right\}$. Consequently, the leakage Hamiltonian is reformulated as: 
\begin{align}
    \mathcal{H}'_{\text{leak}}=&\bigg[\frac{\sqrt{2}}{2}g_{e}e^{i\phi_{e}}\big(|02\rangle\langle11|+|11\rangle\langle20|\big)+\text{H.c.}\bigg]\notag\\
    &-\bigg(\alpha_2+\frac{\Delta_{e}}{2}\bigg)|02\rangle\langle02|+\frac{\Delta_{e}}{2}|11\rangle\langle11|\notag\\
    &+\bigg(\frac{3\Delta_{e}}{2}-\alpha_1\bigg)|20\rangle\langle20|, 
\end{align}
where $g_{e}=2J_1(\beta_1)g_{12}$, $\Delta_{e}=\Delta_t-\eta$, and $\phi_{e}=\eta t+\phi_1-\pi/2$.

Similarly, to mitigate the effects of leakage errors, it is essential to determine the appropriate form of the transformation operator $A_2(\delta_{p_i})$. In the context of a superconducting two-qubit system, the parameter offset can be defined as $\delta_{p_i}=\{\delta_{p_1},~\delta_{p_2},~\dots\}=\{\delta_x',~\delta_y',~\delta_z'\}$. The form of the transformation operator $A_2(\delta_{q,q=x,y,z})$ is defined as
\begin{align}
    A_2(X_2)=\exp[iX_2(\delta'_{q,q=x,y,z})]\red{,}\label{a2}
\end{align}
\begin{align}
    X_2(\delta'_q)=&\delta_z'(-|02\rangle\langle02|+|11\rangle\langle11|+3|20\rangle\langle20|), \notag\\
    &+[(i\delta_y'+\delta_x')(|02\rangle\langle11|+|11\rangle\langle20|)+\text{H.c.}]\red{.}
\end{align}
Evidently, $A(\delta'_q)$ fulfills the Eq. \eqref{cond1}, with parameters $\delta_x'$, $\delta_y'$, and $\delta_z'$ employed to satisfy Eq. \eqref{cond2}. Utilizing the unitary transformation with transformation matrix as Eq. \eqref{a2}, the corrected Hamiltonian is derived as
\begin{align}
    \mathcal{H}_\text{I}^2(t)=&A_2(\delta_q')\mathcal{H}_\text{leak}^2(t)A_2^{\dagger}(\delta_q)+i\dot{A_2}(\delta_q)A_2^{\dagger}(\delta_q')\notag\\
    =&[(1+\delta_{g})\frac{\sqrt{2}}{2}g_{e}e^{i\delta_\phi'}(|02\rangle\langle11|+|11\rangle\langle20|)+\text{H.c.}]\notag\\
    &+\delta_{02}|02\rangle\langle02|\notag+\delta_{20}|20\rangle\langle20|+\frac{\Delta_{e}}{2}|11\rangle\langle11|\\
    &+(\delta_y-i\delta_x)(-\alpha_2'|02\rangle\langle11|+\alpha_1'|11\rangle\langle20|+\text{H.c.}),\label{hi2}
\end{align}
with $\delta_{g}=\sqrt{1+\delta_z^2}-1$, $\delta_\phi'=\arctan(\delta_z)$, $\delta_{02}=-\alpha_2-\Delta_{e}/2+\sqrt{2}\delta_yg_{e}$, $\delta_{20}=-\alpha_1+3\Delta_{e}/2+\sqrt{2}\delta_yg_{e}$, $\alpha_1'=\alpha_1-\Delta_{e}$, $\alpha_2'=\alpha_2+\Delta_{e}$. 
By appropriately offsetting the parameters $\delta_x'$, $\delta_y'$, and $\delta_z'$ in Eq. \eqref{hi2}, one can effectively reduce the amplitudes of the leakage terms $|02\rangle\langle11|$ and $|20\rangle\langle11|$ in the transformed Hamiltonian to values much smaller than the corresponding anharmonic terms, thereby suppressing leakage errors efficiently. It is important to note that in a superconducting two-qubit system, the small and static parameter set offsets correspond to the coupling strength, detuning, and phase offsets in the system Hamiltonian, resulting in $\delta_{g},\,\delta_\Delta',\,\delta_\phi'$. To mitigate the impact of leakage errors, these offsets are applied. Consequently, the offset parameters in superconducting two-qubit quantum systems can be expressed as $\delta_{p_i}=\{\delta_{g},\,\delta_\Delta',\,\delta_\phi'\}$. The offset of coupling strength can be represented by $g_{12}\rightarrow(1+\delta_{g})g_{12}$, which can be equivalently achieved through pulse duration in experimental implementations. Pulse phase and detuning offsets are applied as $\phi_e \rightarrow \phi_e + \delta_\phi'$ and $\Delta_e \rightarrow \Delta_e + \delta_\Delta'$, respectively. It should be noted that, the modulation of $\phi_e$ and $\Delta_e$ is equivalent to the modulation of $\phi_1$ and $\nu_1$.

\begin{figure}
    \centering
    \includegraphics[width=1.0\linewidth]{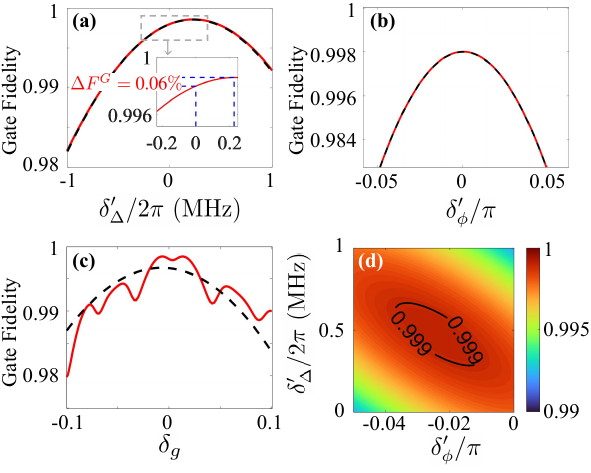}
    \caption{The approximate solutions from calculations (dashed lines) and exact solutions from numerical simulations (solid lines) for gate fidelity as functions of small, static offsets of coupling strength $\delta_\Omega$, phase $\delta_\phi$ and detuning $\delta_\Delta$, respectively, for $i$SWAP gate. $\Delta F^G$ denotes the enhancement in fidelity achieved by introducing offsets, as compared to without offsets.}
    \label{two_offsets}
\end{figure}

We continue to use our approach to construct the leakage-suppression superconducting two-qubit $i$SWAP gate. Additionally, we apply the gate-fidelity calculation formula to determine the relationship between gate fidelity and offsets. The selected qubit parameters are as follows: the coupling strength between qubits is $g_{12}=2\pi\times10~\text{MHz}$; the intrinsic anharmonicities of qubits $T_1$ and $T_2$ are $\alpha_1=2\pi\times220~\text{MHz}$ and $\alpha_2=2\pi\times200~\text{MHz}$, respectively; and the frequency difference between qubits $T_1$ and $T_2$ is $\Delta_1=2\pi\times500~\text{MHz}$. For generality, we select the drive parameter $\beta_1=\varepsilon_1/\nu_1=1.2$. The calculation yields
\begin{subequations} 

\begin{align}
    F^G_{iS}(\delta_\Delta')&\approx-0.011\bigg(\frac{\delta_\Delta'}{2\pi}\bigg)^2+0.005\frac{\delta_\Delta'}{2\pi}+0.998\\
    F^G_{iS}(\delta_\phi')&\approx-2.201\bigg(\frac{\delta_\phi'}{\pi}\bigg)^2+0.0023\frac{\delta_\phi'}{\pi}+0.998\\
    F^G_{iS}(\delta_g)&\approx-1.132\delta_g^2-0.016\delta_g+0.997
\end{align}
\end{subequations}
Figures~\ref{two_offsets}(a)-\ref{two_offsets}(c) illustrate the approximate solution from calculation (dashed lines) and exact solutions from numerical simulations (solid lines) for the gate fidelity function concerning the coupling strength offset $\delta_{g}$, detuning offset $\delta_\Delta'$, and phase offset $\delta_\phi'$ in the $i$SWAP gate. The calculated results align closely with the simulation outcomes. These findings demonstrate that each of the three parameters exerts a certain degree of influence on fidelity, encompassing both positive and negative effects. Consequently, it is essential to strategically optimize the small and static offsets to enhance fidelity positively. 

Similarly, Equation~\eqref{hi2} indicates intrinsic interdependencies among the offset parameters, suggesting that simultaneous adjustment of multiple parameters is required for optimal performance. As clearly illustrated in Fig.~\ref{two_offsets}(d), jointly offsetting the detuning $\delta'_\Delta$ and the phase parameter $\delta'_\phi$ enables further suppression of leakage errors. We then extend the analysis to the simultaneous tuning of three parameters. Choosing coupling strength offset $\delta_g = 0.004$, detuning offset $\delta'_\Delta = 2\pi\times0.52~\text{MHz}$, and phase offset $\delta'_\phi = -0.026\pi$ improves the gate fidelity from $99.80\%$ to $99.94\%$. Notably, all offsets remain modest in magnitude and well within experimentally accessible ranges, ensuring that the proposed procedure introduces no additional experimental complexity. Similarly, as we show in Appendix \ref{AppA}, the scheme still maintains a significant leakage suppression effect even at lower experimental precision.

\begin{figure}
    \centering
    \includegraphics[width=1.0\linewidth]{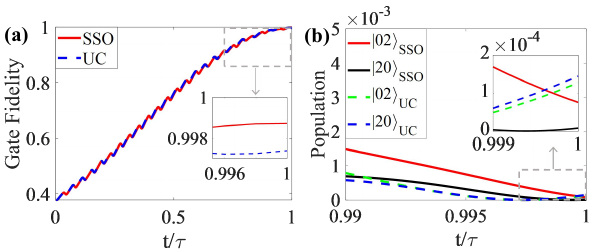}
    \caption{Figure (a) compares the fidelity of the $i$SWAP gate using optimized parameters under the small, static offsets (SSO) scheme and uncorrected (UC), taking into account decoherence effects. Figure (b) shows the state population for $i$SWAP gate which the initial state is $|01\rangle$, similarly considering decoherence effects.}
    \label{two_fidelity_population}
\end{figure}

To comprehensively evaluate the effectiveness of this scheme in specific physical implementations, we consider the impact of decoherence. By solving the quantum master equation, we obtain the density operator $\rho_{q2}$ for the two-qubit system under the influence of decoherence. We define the fidelity as $F_2^G=\frac{1}{4\pi^2}\int_{0}^{2\pi}\int_{0}^{2\pi}\langle\varPhi_{\text{q}2}|\rho_{\text{q}2}|\varPhi_{\text{q}2}\rangle d\vartheta_1d\vartheta_2$ \cite{singlefg1,singlefg2}. Here, $|\varPhi_{\text{q}2}\rangle=U_0'|\varPhi_2\rangle$ represents the ideal final state of the general initial state of two logical qubits, $|\varPhi_2\rangle=(\cos\vartheta_1|0\rangle+\sin\vartheta_1|1\rangle)\otimes(\cos\vartheta_2|0\rangle+\sin\vartheta_2|1\rangle)$, where $U_0'$ is the ideal evolution operator of the $i$SWAP gate. The $i$SWAP gate produces the ideal final state $|\varPhi_{fiS}\rangle=\cos\vartheta_1\cos\vartheta_2|00\rangle+i\cos\vartheta_1\sin\vartheta_2|10\rangle+i\sin\vartheta_1\cos\vartheta_2|01\rangle+\sin\vartheta_1\sin\vartheta_2|11\rangle$. We select the decoherence parameters as $\kappa_1=\kappa_\varphi=2\pi\times2~\text{KHz}$. As illustrated in Fig.~\ref{two_fidelity_population}(a), considering decoherence, the fidelity of the $i$SWAP gate increases from $99.74\%$ to $99.87\%$ when the small and static offsets to system parameters are optimally selected. Moreover, our scheme demonstrates robustness against calibration errors, as clearly illustrated in Appendix \ref{AppA}, which indicates that the scheme can maintain high fidelity across a broad range of error values. Additionally, we assume the qubit is initially in the state $|01\rangle$, and the ideal final state produced by the $i$SWAP is $|\psi_\text{iS}\rangle=i|10\rangle$. As shown in Fig.~\ref{two_fidelity_population}(b), the suppression of leakage via small, static offsets applied to tunable system parameters can also be evaluated by monitoring the corresponding state populations.

\section{EXTENDED APPLICABILITY OF THE SCHEME}

We have demonstrated the effective suppression of leakage errors in single-qubit systems and successfully extended this approach to two-qubit gates. We will further adapt this method to be compatible with perfect state transfer in multi-level system and optimization control techniques, thereby enhancing the overall performance of complex quantum circuits while continuing to suppress leakage errors.

\subsection{Perfect State Transfer in Multi-Level System}

This subsection further investigates the suppression of leakage in stimulated Raman adiabatic transitions \cite{RAP1,RAP2,RAP3,RAP4} within multi-level systems. Using superconducting transmon-type qubits \cite{PulseOP2,LL1,LL2,LL4} as an illustrative example, we aim to demonstrate that small, static parameter offsets can effectively mitigate leakage, thereby enabling perfect state transfer \cite{ST1,ST2}.

To facilitate state transfer between $|0\rangle$ and $|2\rangle$, while accounting for leakage errors, we employ a ladder configuration characterized by the energy level structure $|0\rangle \rightarrow |1\rangle \rightarrow |2\rangle \rightarrow |3\rangle$, as illustrated in Fig.~\ref{F1}(c). In this configuration, $|0\rangle$ and $|2\rangle$ serve as the target levels, where $|1\rangle$ serve as the auxiliary level. Two independent microwave driving fields are utilized: microwave $D_a$ is responsible for the $|0\rangle \leftrightarrow |1\rangle$ transition, featuring a pulse waveform $\Omega_{01}(t)$ and phase $\phi_{01}(t)$. The application of $D_a$ induces non-ideal interactions between $|1\rangle \leftrightarrow |2\rangle$ and $|2\rangle \leftrightarrow |3\rangle$, resulting in leakage. Similarly, microwave $D_b$ drives the $|1\rangle \leftrightarrow |2\rangle$ transition, characterized by a pulse waveform $\Omega_{12}(t)$ and phase $\phi_{12}(t)$. The influence of $D_b$ leads to non-ideal interactions for $|0\rangle \leftrightarrow |1\rangle$ and $|2\rangle \leftrightarrow |3\rangle$, which also resulting in leakage.

Under the interaction picture, within the rotating-wave approximation, the system is described by
\begin{align}
    \mathcal{H}^3(t)=\mathcal{H}^3_\text{targ}(t)+\mathcal{H}_\text{leak}^3(t)
\end{align}
Here, $\mathcal{H}_\text{targ}^3(t)$ denotes the target Hamiltonian of the system, given by 
\begin{align}\label{Htarg}
    \mathcal{H}_{\text{targ}}^3(t)=&\Omega_{01}(t)e^{i(\phi_{01}-\Delta_{01} t)}|0\rangle\langle 1|\notag\\
    &+\Omega_{12}(t)e^{i(\phi_{12}-\Delta_{12} t)}|1 \rangle\langle 2|+{\rm H.c.}
\end{align}
where $\Delta_{01}$ and $\Delta_{12}$ represent the frequency differences between the qubit frequencies and the target driving microwaves $D_a$ and $D_b$, respectively. 

An eigenstate of $\mathcal{H}^3_\text{targ}(t)$, denoted as $|d\rangle = \cos(\theta/2)|0\rangle + \sin(\theta/2)e^{i\phi}|2\rangle$, is identified as a dark state, where $\phi = \phi_{01} - \phi_{12} + \pi$, $\tan\theta = \Omega_{01}(t)/\Omega_{12}(t)$, and $\Omega(t) = \sqrt{\Omega_{01}(t)^{2} + \Omega_{12}(t)^{2}}$. This dark state $|d\rangle$ is decoupled from the system's dynamical evolution, allowing the system's dynamics to be described by the coupling interaction between the states $|b\rangle = \sin(\theta/2)e^{i\phi}|0\rangle - \cos(\theta/2)|2\rangle$ and $|1\rangle$. Furthermore, for any $j, i \in \{b, d\}$ satisfying $\langle j|\mathcal{H}_{\text{targ}}^3(t)|i\rangle = 0$, no transition occurs between the states $|d\rangle$ and $|b\rangle$ during evolution, thereby fulfilling the parallel transport condition, with the dynamic phase of both $|d\rangle$ and $|b\rangle$ being zero. Consequently, the coherent population transfer from state $|0\rangle$ to $|2\rangle$ can be achieved by setting the evolution angle $\theta$ and utilizing the auxiliary level $|1\rangle$.

Nevertheless, in practical scenarios, leakage errors are often inevitable. Here, we define $\mathcal{H}^3_\text{leak}(t)$ as the leakage Hamiltonian of the system, explicitly represented by: 
\begin{widetext}
\begin{align}\label{Hleak}
&\mathcal{H}^3_\text{leak}(t)=\mathcal{H}_{\text{leak01}}^3(t)+\mathcal{H}_{\text{leak12}}^3(t),\notag\\
    &\mathcal{H}_{\text{leak01}}^3(t)= {\Omega_{01}(t)}\left[\sqrt{2}e^{i[\phi_{01}-(\Delta_{01}-\alpha)t]}|1\rangle\langle2| + \sqrt{3}e^{i[\phi_{01}-(\Delta_{01}-2\alpha)t]}|2\rangle\langle3| + \text{H.c.} \right],\notag\\
&\mathcal{H}_{\text{leak12}}^3(t)={\Omega_{12}(t)} \left[ \frac{1}{\sqrt{2}}e^{i[\phi_{12}-(\Delta_{12}-\alpha)t]}|0\rangle\langle1|+  \frac{\sqrt{3}}{\sqrt{2}}e^{i[\phi_{12}-(\Delta_{12}-2\alpha)t]}|2\rangle\langle3| + \text{H.c.} \right].
\end{align}
\end{widetext}

Two independent pulse drives are shown in Fig.~\ref{F1}(d): Pulse $D_a$ drives the $|0\rangle \leftrightarrow |1\rangle$ transition to produce the target transition, while also causing non-ideal coupling terms $\mathcal{H}_{\text{leak01}}^3(t)$ for $|1\rangle \leftrightarrow |2\rangle$ and $|2\rangle \leftrightarrow |3\rangle$. Similarly, pulse $D_b$ drives the $|1\rangle \leftrightarrow |2\rangle$ transition to produce the target transition, while also causing non-ideal coupling terms $\mathcal{H}_{\text{leak12}}^3(t)$ for $|0\rangle \leftrightarrow |1\rangle$ and $|2\rangle \leftrightarrow |3\rangle$. As previously discussed, we use a $sin$-type pulse waveform $\Omega(t) = \Omega_m \sin(\pi t/\tau)$ as an example, where $\Omega_m$ represents the peak value of the pulse and $\tau$ is the pulse duration. The driving forms of the two independent pulses are given by $\Omega_{01} = \Omega(t) \sin(\theta/2)$ and $\Omega_{12} = \Omega(t) \cos(\theta/2)$, respectively.

We focus on utilizing the auxiliary level $|1\rangle$ to facilitate state transfer between the target levels $|0\rangle$ and $|2\rangle$. The general channel of a three-level system is parametrically defined as \cite{RAP1}
\begin{align}
    |\Phi_{\rm{RAP}}\rangle =& \cos\gamma\,\cos\theta|0\rangle + e^{i\phi_{01}} \sin\gamma\,|1\rangle\notag \\
    &- e^{i\phi_{12}}\,\cos\gamma\,\sin\theta|2\rangle,
\end{align}
where $\gamma(t)$ represents the global phase. To achieve state transfer from $|0\rangle$ to $|2\rangle$, we impose the following boundary conditions: at times $t=0$ and $t=T$, $\gamma(0)=\gamma(T)=0$, $\theta(0)=0$, and $\theta(T)=\pi/2$. When the cyclic evolution condition $\int_{0}^{T}\Omega(t)dt=\pi$ is satisfied, specific single-qubit gates can be realized by selecting different values of $\theta$ and/or $\phi$. 
\begin{figure}
    \centering
    \includegraphics[width=1.0\linewidth]{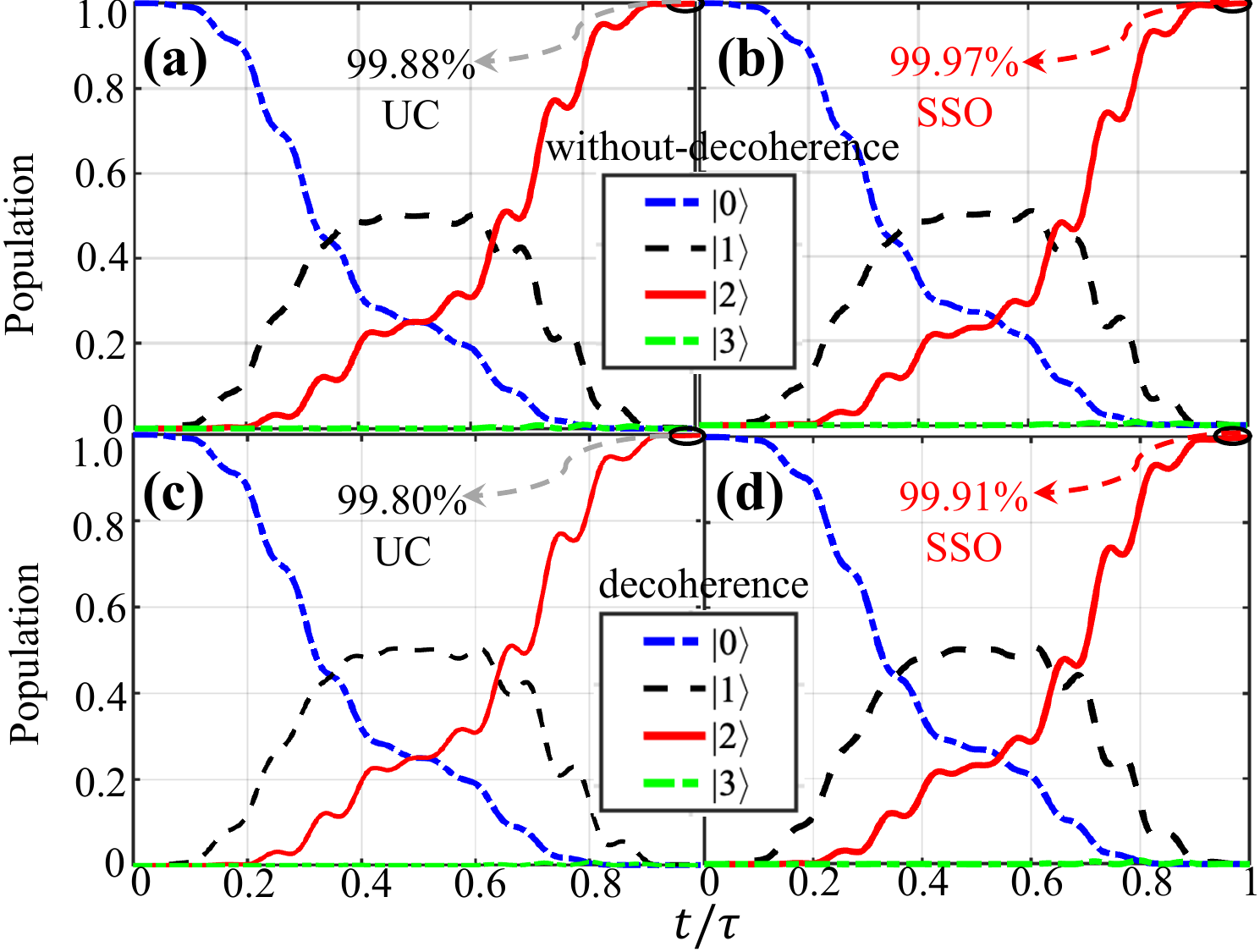}
    \caption{State fidelity and population of each energy level. (a) Considering leakage errors but without decoherence, the state transfer process under uncorrected (UC). (b) Considering leakage errors but without decoherence, small, static offsets (SSO) is incorporated to actively suppress leakage errors in state transfer. (c) When considering leakage errors and incorporating decoherence simultaneously, the state transfer process without correction. (d) Considering both leakage errors and decoherence simultaneously, SSO is introduced during state transfer to actively suppress leakage errors.  }
    \label{F6}
\end{figure}
In the quantum state \(\left| \Phi_{RAP} \right\rangle\), the populations of the three levels \(|0\rangle\), \(|1\rangle\) and \(|2\rangle\) are determined by the squared magnitudes of their probability amplitudes: $P_0(t) = \cos^2\gamma(t) \cos^2\theta(t)$, $P_1(t) = \sin^2\gamma(t)$, and $P_2(t) = \cos^2\gamma(t) \sin^2\theta(t)$, respectively, while satisfying the normalization condition. Under the boundary conditions \(\gamma(0)=\gamma(T)=0\), \(\theta(0)=0\), and \(\theta(T)=\pi/2\), the system ideally transitions from the initial state \(|0\rangle\) to the state \(|2\rangle\) without leakage or decoherence. Throughout the evolution, the population of the intermediate state \(|1\rangle\) is solely determined by the parameter \(\gamma(t)\), while the populations of \(|0\rangle\) and \(|2\rangle\) are influenced by both \(\gamma(t)\) and \(\theta(t)\). The population ratio between \(|0\rangle\) and \(|2\rangle\) is given by \(\theta(t)\) as \(P_0/P_2 = \cot^2\theta(t)\). However, during the actual application of pulses, unavoidable presence of leakage terms \(\mathcal{H}_\text{leak}^3(t)\) can introduce leakage errors across different energy levels. To address this, we will actively apply small and static offsets to the Hamiltonian during multi-level state transfer to suppress passive leakage errors.

We assess the effectiveness of small, static offsets in multi-level state transfer by examining state fidelity and population through numerical simulations. To account for the impact of decoherence, we solve the quantum master equation to obtain the corrected density operator \(\rho_m\) of the multi-level system. A comprehensive evaluation is then conducted using the state fidelity \cite{singlefg1,singlefg2} defined by $F^G=\langle\psi_{fN2}|\rho_m|\psi_{fN2}\rangle$. Assuming the qubit initially resides in the state \(|\psi_2\rangle=|0\rangle\), the ideal state transfer results in the final state \(|\psi_{fN2}\rangle=|2\rangle\). Consistent with the previously mentioned experimental parameters, we set \(\kappa_1=\kappa_\varphi=2\pi\times2~\text{kHz}\), \(\Omega_m=2\pi\times30~\text{MHz}\), and \(\alpha=2\pi\times220~\text{MHz}\). In the absence of decoherence with optimal offset parameters \(\delta_\Omega=-0.026,~\delta_\Delta=2\pi\times1.04~\text{MHz},~\delta_\phi=-0.03\pi\), as shown in Figs.~\ref{F6}(a)-\ref{F6}(b), the state fidelity in the multi-level state transfer system improves from $99.88\%$ to $99.97\%$. Similarly, when decoherence is taken into account with optimal offset parameters, as shown in Figs.~\ref{F6}(c)-\ref{F6}(d), the state fidelity improves from $99.80\%$ to $99.91\%$. It is evident that introducing small, static
offsets into multi-level state transfer can effectively suppress leakage errors.

\subsection{Crosstalk Suppression via Optimal Control}

\begin{figure}
    \centering
    \includegraphics[width=1.0\linewidth]{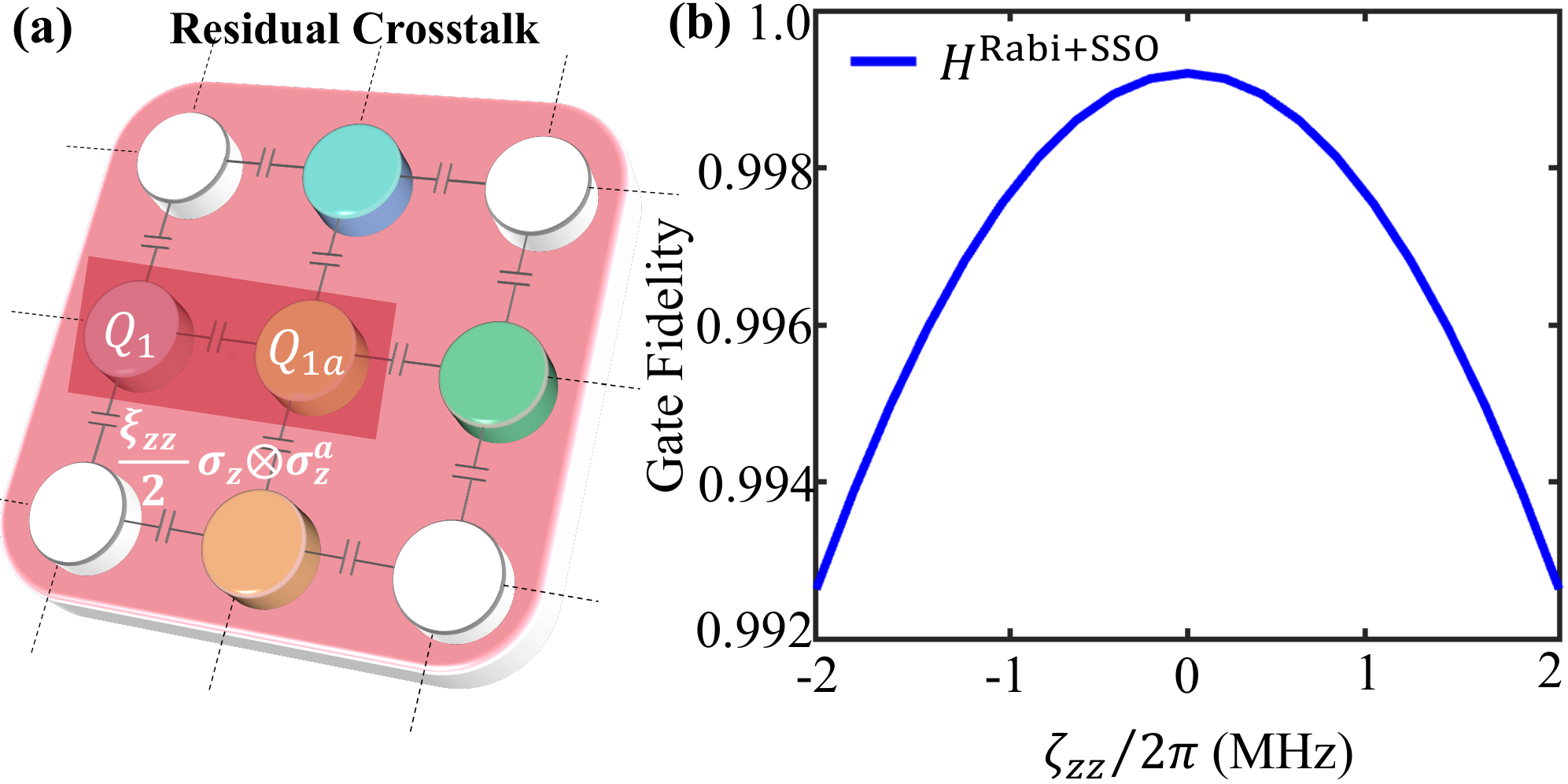}
    \caption{(a) Schematic diagram of a two-dimensional superconducting qubit lattice, where there is residual ZZ crosstalk errors between the target control qubit $Q_1$ and its neighboring spectator qubit $Q_{1a}$. (b) The fidelity of the Rabi $H$ gate under residual ZZ crosstalk errors with active suppression of leakage errors using small static offset (SSO) integration. It can be noted that the gate fidelity is highly sensitive to crosstalk errors, leading to a sharp decline in performance.}
    \label{F7}
\end{figure}

In this section, we introduce a scheme for implementing robust quantum gates using optimal control techniques \cite{OCT1,OCT2,OCT3,OCT4,OCT5}, which can be combined with small and static offsets to jointly suppress residual crosstalk and leakage errors. This approach enhances the performance of complex quantum circuits, providing a practical and efficient path to achieving robust, high-fidelity quantum operations. Notably, crosstalk, a critical error in superconducting quantum circuits, inevitably leads to leakage accumulation, significantly impacting gate fidelity. It is important to emphasize that due to the inability to spatially decouple the dynamic behavior of neighboring qubits, residual weak entanglement effects persist, particularly residual ZZ crosstalk \cite{ZZ1,ZZ2,ZZ3,ZZ4,ZZ5,ZZ6}.

Next, we take a two-dimensional qubit lattice structure as shown in Fig.~\ref{F7}(a) as an example, where there is an effective ZZ interaction with strength \(\zeta_{zz}\) between the target qubit \(Q_1\) and the neighboring spectator qubit \(Q_{1a}\). This interaction is described by \(V_{zz} = \frac{\zeta_{zz}}{2}\sigma_z \otimes \sigma^{a}z\), representing residual ZZ crosstalk. Here, \(\sigma_z\) and \(\sigma^{a}_z\) are the Pauli Z operators for \(Q_1\) and \(Q_{1a}\), respectively. We utilize the previously proposed geometric trajectory correction \cite{GTC1}, with the Hamiltonian form as in Eq. (\ref{Htarg}) , to achieve suppression of residual ZZ crosstalk errors. It is noted that the presence of leakage terms in the form of Eq. (\ref{Hleak}) inevitably hinders the achievement of high gate fidelity. Additionally, the presence of these leakage terms significantly undermines the advantage of using geometric trajectory correction to suppress ZZ crosstalk. Therefore, we employ the proposed method of small and static offsets to maximize the suppression of leakage while ensuring the suppression of ZZ crosstalk by geometric trajectory correction, thereby demonstrating the compatibility of multiple schemes. Based on the gate fidelity \cite{singlefg1,singlefg2} construction method in Section \ref{GF}, we also numerically simulate the Hadamard gate fidelity of the Rabi scheme in the presence of residual ZZ crosstalk errors, as shown in Fig.~\ref{F7}(b). It is evident that the results show a sharp decline. Based on this, in the physical implementation of superconducting circuits, enhancing the overall performance of complex quantum circuits through the compatibility of multiple schemes is crucial for achieving robust, high-fidelity quantum operations and gate fidelity.

Next, we use geometric trajectory correction techniques to actively suppress ZZ crosstalk errors. The form of the Hamiltonian control parameters corresponding to the geometric trajectory correction is as follows \cite{GTC1}:
\begin{subequations} 
\begin{align} 
\!\!\!\!\!&t\in[0,\tau_1]:
\int^{\tau_1}_0\Omega(t)\textrm{d}t=\chi_0-\chi_1, \ \
\phi(t)=\xi_0-\frac{\pi}{2},
\\
\!\!\!\!\!\begin{split} 
t\in[\tau_1,\tau_2]:
\int^{\tau_2}_{\tau_1}\Omega(t)\textrm{d}t=(\xi_2-\xi_0)\sin\chi_1 \cos\chi_1,
\\
\!\!\!\phi(t)=\xi_0+\pi+\frac{\int \Omega(t)\textrm{d}t}{\sin\chi_1 \cos\chi_1},
\end{split}
\\
\!\!\!\!\!&t\in[\tau_2,\tau_3]:
\int^{\tau_3}_{\tau_2}\Omega(t)\textrm{d}t=\chi_3-\chi_1,
\phi(t)=\xi_2+\frac{\pi}{2},
\\
\!\!\!\!\!\begin{split}
t\in[\tau_3,\tau_4]:
\int^{\tau_4}_{\tau_3}\Omega(t)\textrm{d}t=(\xi_0-\xi_2)\sin\chi_2 \cos\chi_2,
\\
\!\phi(t)=\xi_2+\pi+\frac{\int \Omega(t)\textrm{d}t}{\sin\chi_2 \cos\chi_2},
\end{split}
\\
\!\!\!\!\!&t\in[\tau_4,\tau]:
\int^{\tau}_{\tau_4}\Omega(t)\textrm{d}t=\chi_3-\chi_0,\ \
\phi(t)=\xi_0-\frac{\pi}{2},
\end{align}
\end{subequations}
The detuning for each time fragment are \(\Delta(t) = 0, -\Omega(t) \tan \chi_1, 0, -\Omega(t) \tan \chi_3\), and \(0\). Here, \(\chi(t)\) and \(\xi(t)\) correspond to the polar and azimuthal angles of the two-dimensional dressed state on the Bloch sphere, as depicted in the Hamiltonian in Eq. (\ref{Htarg}).

The relationship between $\chi(t)$ and $\xi(t)$ and the control of the Hamiltonian $\mathcal{H}_\text{target}^{1}(t)$ can be established through two evolution states $|\Psi_{1,2}(t)\rangle$ that satisfy the Schr\"{o}dinger equation $i\frac{\partial}{\partial t}|\Psi_k(t)\rangle = \mathcal{H}_\text{target}^{1}(t)|\Psi_k(t)\rangle$. The specific forms of these evolution states are: 
\begin{align}
|\Psi_1(t)\rangle &= e^{i f_1(t)} \left[ \cos \frac{\chi(t)}{2} |0\rangle + \sin \frac{\chi(t)}{2} e^{i \xi(t)} |1\rangle \right]\\
|\Psi_2(t)\rangle &= e^{i f_2(t)} \left[ \sin \frac{\chi(t)}{2} e^{-i \xi(t)} |0\rangle - \cos \frac{\chi(t)}{2} |1\rangle \right]
\end{align}
Here, $f_{1,2}(t)$ are the accumulated global phases, satisfying $f_{1}(0)=-f_2(0) = 0$. From this, we derive the relationship between the state evolution parameters and the Hamiltonian control parameters: $\dot{\chi}(t)= \Omega(t) \sin[\phi(t) - \xi(t)]$ and $\dot{\xi}(t) = -\Delta(t) - \Omega(t) \cot\chi(t) \cos[\phi(t) - \xi(t)]$. Additionally, the evolution operator over the complete evolution time $\tau$ is given by Eq. (\ref{Utau}). The parameters \({\chi_0, \xi_0, \gamma_g}\) clearly correspond to specific gate parameters, where \(\gamma_g = (\xi_2 - \xi_0)[\cos\chi_1 - \cos\chi_3]/2\) denotes the geometric phase associated with the overall geometric trajectory correction process.
\begin{widetext}
\begin{align}\label{Utau}
U_(\chi_0,\xi_0,\gamma_g) =\begin{pmatrix}\cos{\gamma_g}+i\,\sin{\gamma_g}\cos\chi_0&i\,\sin{\gamma_g}\,\sin\chi_0 e^{-i\xi_0}\\i\,\sin{\gamma_g}\,\sin\chi_0 e^{i\xi_0}&\cos{\gamma_g}-i\,\sin{\gamma_g}\cos\chi_0\end{pmatrix}.
\end{align}
\end{widetext}

\begin{figure}
    \centering
    \includegraphics[width=1.0\linewidth]{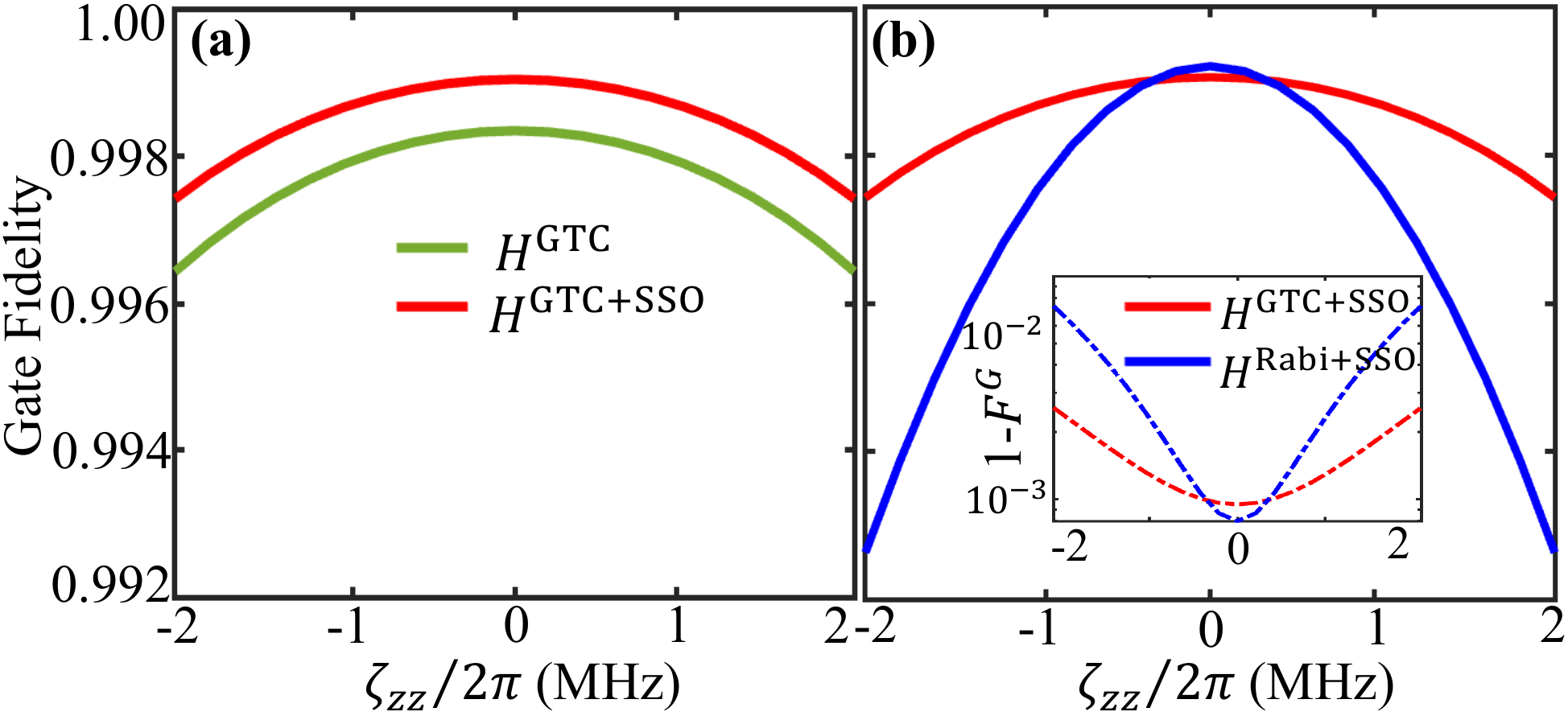}
    \caption{(a) Gate fidelity $F^G$ under residual ZZ crosstalk errors achieved by the proposed geometric trajectory correction (GTC) method: without SSO (green line) and with SSO applied (red line). (b) Adopting a multi‑scheme compatible approach, where geometric trajectory correction (red line) coexists with SSO. We compare its active suppression of ZZ crosstalk errors in terms of gate fidelity $F^G$ and gate sensitivity $1-F^G$ with the previously described Rabi scheme, which only uses SSO (blue line).}
    \label{F8}
\end{figure}

By setting the starting point \(\chi_0\) and \(\xi_0\) of the cyclic trajectory, along with the geometric phase \(\gamma_g\), we can implement arbitrary geometric gates. This approach demonstrates that the additional parameters offer ample optimization freedom to address ZZ crosstalk. We use the geometric Hadamard gate \(H^\text{G}\) as an example for gate construction and optimization analysis. Initially, we determine the basic form of the geometric gate \(H^\text{G}\) by fixing the parameters \(\{\chi_0, \xi_0, \gamma_g\} = \{\pi/4, 0, 3\pi/2\}\). In this configuration, the free optimization parameters of the trajectory are \(\{\chi_1, \chi_3\}\), while the remaining evolution parameters are linked to \(\gamma_g\), collectively completing the gate construction. The optimization ranges for the two adjustable parameters are \(\chi_1 \in [0, \pi/4]\) and \(\chi_3 \in [\pi/4, \pi]\). Then, we optimize the trajectory within the ranges to determine the optimal design to suppress ZZ crosstalk errors. Through numerical simulation, we select \(\{\chi_1, \chi_3\} = \{0, 0.7\pi\}\) as the optimal trajectory parameters for crosstalk suppression. In this case, the five-segment path based on geometric trajectory correction is simplified into a four-segment triangular cyclic trajectory \cite{GTC1}.

One of the primary limitations of previous geometric schemes in terms of control robustness is their restricted range of geometric evolution trajectories, which cannot actively avoid segments severely impacted by system errors. To address this, we introduce a sufficient number of evolution parameters through geometric trajectory correction to effectively reduce sensitivity to system errors. The optimal trajectory parameters \(\{\chi_1, \chi_3\}\) correspond to the trajectory that best suppresses residual crosstalk, thereby minimizing the impact of system errors. Furthermore, we achieved active compatibility with small, static offsets, significantly suppressing leakage errors and further enhancing gate fidelity.

Similarly, in superconducting circuits, using this compatible scheme with the \(H^\text{G}\) gate as an example, we consider the presence of the same leakage terms as previously discussed. We evaluate the effect of this compatible scheme by observing gate fidelity through numerical simulations. Under the influence of decoherence, we perform a comprehensive evaluation of \(H^\text{G}\) using the single-qubit gate fidelity construction described previous, maintaining the same experimental parameters as before: \(\kappa_1=\kappa_\varphi=2\pi\times2~\text{kHz}\), \(\Omega_m=2\pi\times30~\text{MHz}\), \(\alpha=2\pi\times220~\text{MHz}\). After incorporating small and static offsets into the geometric trajectory correction, the fidelity of the \(H^\text{G}\) gate improves from $99.80\%$ to $99.91\%$, as shown in Fig.~\ref{F8}(a). Next, by evaluating the fidelity of the $H^G$ gate, we compare this compatible scheme with the scheme based on the Rabi process, which works by actively suppressing ZZ crosstalk errors through the introduction of only small, static offsets. As shown in Fig.~\ref{F8}(b), we observe that the gate fidelity of the single SSO scheme becomes extremely sensitive to ZZ crosstalk errors, leading to fidelity degradation. In contrast, the multi-scheme compatible approach demonstrates strong active suppression against both crosstalk and leakage errors, exhibiting significant advantages. This indicates that when small and static offsets are combined with optimal control techniques, our method can collaboratively suppress leakage errors and residual crosstalk errors, thereby enhancing the overall performance of complex quantum circuits and achieving high gate fidelity. This paves a practical and efficient path for achieving robust, high-fidelity quantum operations.\\

\section{Conclusion}

In conclusion, we have presented a general strategy for suppressing system leakage errors by applying small, static offsets to tunable system parameters, without modifying the original control framework or incurring additional time overhead. This approach effectively mitigates leakage's detrimental impact on quantum control and remains fully compatible with subsequent optimal control techniques. Numerical validation on superconducting quantum circuits demonstrates effective leakage suppression, enabling high-fidelity single-qubit gates, precise control of two-qubit interactions, and perfect state transfer in multi-level systems. Moreover, when integrated with optimal control, it allows for the cooperative suppression of both leakage errors and residual crosstalk.

\acknowledgments

T. Lin and Z. H. Qin contributed equally to this work. This work was supported by the National Natural Science Foundation of China (Grants No. 12305019 and 92576110), and the Guangdong Provincial Quantum Science Strategic Initiative (Grant No. GDZX2203001).

\section*{DATA AVAILABILITY}

The data that support the findings of this article are notpublicly available. The data are available from the authors upon reasonable request.

\begin{appendices}
\section{ANALYSIS OF TOLERANCE TO OFFSET CALIBRATION}\label{AppA}

In this section, we quantitatively analyze the tolerance of error correction based on the small, static offsets (SSO) scheme, focusing on scenarios involving varying levels of parameter precision and deviations in offset parameters. It is important to emphasize that our approach does not alter the control framework, maintaining consistent sensitivity to errors as before the modification. The correction process also preserves robustness. Furthermore, our scheme imposes no restrictions on pulse waveforms. All our offset parameters achieve precision within the range currently achievable in advanced experiments. Nevertheless, a dedicated analysis of the tolerance and accuracy of the optimized offset parameters is crucial for future experimental implementation.


\begin{figure*}
    \centering
    \includegraphics[width=0.7\linewidth]{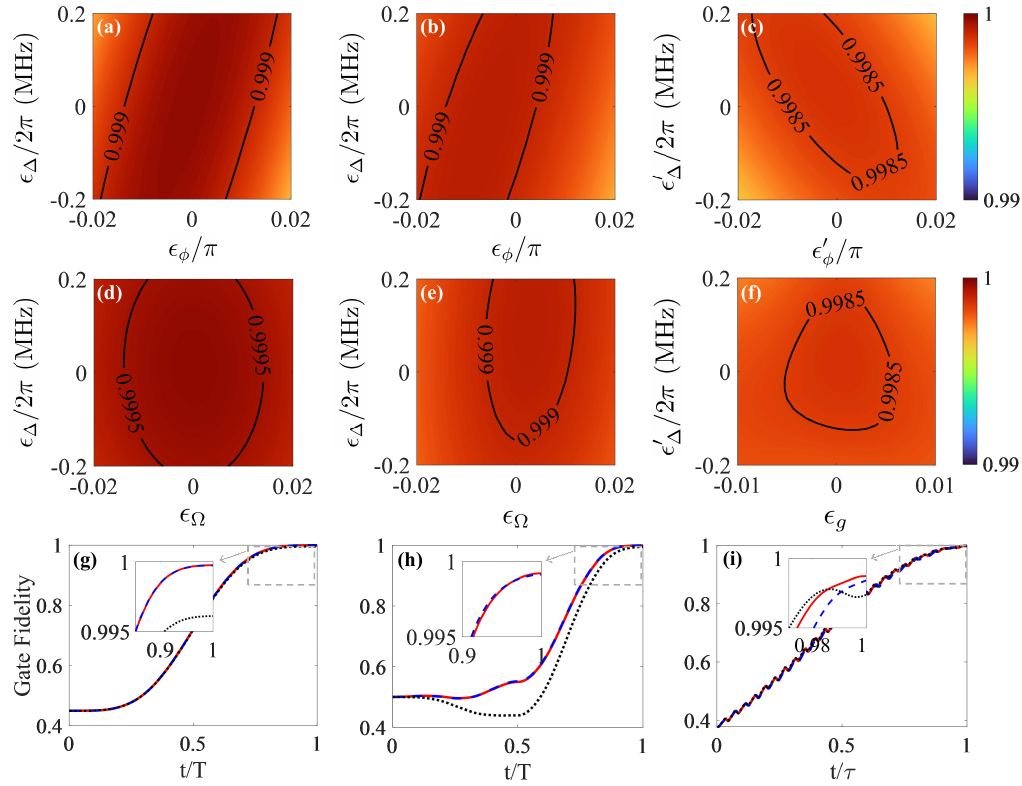}
    \caption{Gate fidelity under optimal small, static offsets of tunable parameters, including decoherence. (a)-(c) show the gate fidelity versus phase and detuning calibration errors for the NOT, Hadamard, and $i$SWAP gates, respectively. (d)-(f) show the gate fidelity versus coupling strength and detuning calibration errors for the NOT, Hadamard, and $i$SWAP gates, respectively. (g)-(i) display the gate fidelity for the NOT, Hadamard, and $i$SWAP gates under different offset precisions. The solid red line indicates that the precision for both coupling strength offset and detuning offset is $2\pi\times0.01\text{MHz}$, while the phase offset precision is $0.001\pi$. The dashed blue line indicates that the precision for both coupling strength offset and detuning offset is $2\pi\times0.1\text{MHz}$, while the phase offset precision is $0.01\pi$. The dashed black line represents the gate fidelity without any applied corrections.}
    \label{F9}
\end{figure*}

First, we analyze the robustness of single-qubit (using the NOT gate and Hadamard gate as examples) and two-qubit gates (using the $i$SWAP gate as an example) separately. For single-qubit gates, we define the experimental errors in coupling strength, detuning, and phase as $\{\Omega_m\rightarrow(1+\delta_{\Omega}+\epsilon_{\Omega})\Omega_m,~ \Delta\rightarrow\Delta+\delta_{\Delta}+\epsilon_{\Delta},~\phi\rightarrow\phi+\delta_{\phi}+\epsilon_{\phi}\}$, where $\epsilon_{p}=\{\epsilon_{\Omega},~ \epsilon_{\Delta}, ~\epsilon_{\phi}\}$ represent the magnitudes of the deviations (calibration errors) in the tunable parameters, $\delta_{p}=\{\delta_{\Omega},~ \delta_{\Delta},~ \delta_{\phi}\}$ represents the offsets of coupling strength, detuning, and phase, respectively. The ranges for these calibration errors $\{\epsilon_{\Omega}, ~\epsilon_{\Delta}, ~\epsilon_{\phi}\}$ are set within the possible ranges of experimental deviations: $\epsilon_{\Omega}$ in the range of [-0.02, 0.02] (which corresponds to the deviation of $\Omega_m$ in the range of $[-2\pi\times0.6 ~\text{MHz},\ 2\pi\times0.6~\text{MHz}]$), and $\epsilon_\Delta$ in the range of $[-2\pi\times0.2 ~\text{MHz},\ 2\pi\times0.2~\text{MHz}]$, and $\epsilon_\phi$ in the range of $[-0.02\pi, 0.02\pi]$. We evaluate the gate fidelity under decoherence effects, Figs. \ref{F9}(a)(d) and Figs. \ref{F9}(b)(e) demonstrate that both the NOT gate and the Hadamard gate maintain high fidelity, exceeding 99.9$\%$ across a broad parameter range within the specified error bounds, indicating that the scheme exhibits strong robustness even in the presence of calibration errors.

Similarly, in the investigation of gate performance under parameter deviations for two-qubit gate, we define the coupling strength, detuning, and phase with parameter offsets and calibration errors as follows: \{\( g_{12} \to (1 + \delta_g + \epsilon_g)g_{12} \), \( \Delta_e \to \Delta_e + \delta'_\Delta + \epsilon'_\Delta \), \(\phi_e \to \phi_e + \delta'_\phi + \epsilon'_\phi\}\). The ranges for these calibration errors $\{\epsilon_{g},~\epsilon'_{\Delta},~ \epsilon'_{\phi}\}$ are also determined based on classical experimental parameters: $\epsilon_{g}$ in the range of [-0.01, 0.01] (which corresponds to the deviation of $g_{12}$ in the range of $[-2\pi\times0.1 ~\text{MHz},\ 2\pi\times0.1~\text{MHz}]$), and $\epsilon'_\Delta$ in the range of $[-2\pi\times0.2 ~\text{MHz},\ 2\pi\times0.2~\text{MHz}]$, and $\epsilon'_\phi$ in the range of $[-0.02\pi, 0.02\pi]$. We evaluate the $i$SWAP gate fidelity while accounting for decoherence effects. The results presented in Figs. \ref{F9}(c)(f) demonstrate that the $i$SWAP gate also maintains high fidelity, exceeding 99.85\% across a broad parameter range within the specified error bounds.

Furthermore, while the parameter precision used in the main text is within the acceptable range for experiments, we also analyzed the effectiveness of our scheme under reduced parameter control precision. Considering the decoherence effect and setting the precision range in the original (the precision for coupling strength offset and detuning offset is $2\pi\times0.01~\text{MHz}$, while the precision for phase offset is $0.001\pi$), the fidelity of the NOT gate is improved from 99.61$\%$ to 99.98$\%$. The fidelity of Hadamard gates has been enhanced from 99.46$\%$ to 99.92$\%$, and that of $i$SWAP gates has been increased from 99.74$\%$ to 99.87$\%$. The fidelity of the NOT gate is further set to 99.97$\%$, that of the Hadamard gate to 99.91$\%$, and that of the $i$SWAP gate to 99.84$\%$ by further setting the precision reduction (the precision for coupling strength offset and detuning offset is $2\pi\times0.1~\text{MHz}$, while the precision for phase offset is $0.01\pi$). It can be seen that within the acceptable precision range of the experiment, our scheme can have a very good suppression effect on leakage errors. Even with reduced parameter precision, the leakage suppression effect remains significant, with fidelities exceeding 99.9\% for single-qubit gates and 99.8\% for the two-qubit gate, as shown in Figs. \ref{F10}(g)–\ref{F10}(i).

In summary, our scheme shows good robustness in the presence of calibration errors within a reasonable range, and can still maintain high fidelity under conditions lower than the current experimental precision, which has a good prospect for experimental implementation.

\section{QUANTITATIVE CONTRAST WITH DRAG } \label{AppB}
\begin{figure}[htbp]
    \centering
    \includegraphics[width=1\linewidth]{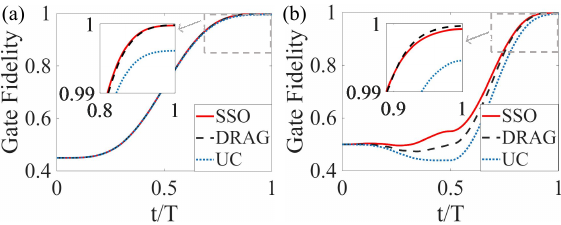}
    \caption{Gate fidelity of two schemes under decoherence effects for (a) the NOT gate and (b) the Hadamard gate. The solid red line denotes the application of small, static offsets (SSO) to the tunable parameters of the Hamiltonian, while the black dashed line represents the DRAG scheme. The blue dashed line indicates the scenario where no corrections are applied.}
    \label{F10}
\end{figure}

In this section, we demonstrate the contributions of our research by comparing our SSO scheme with state-of-the-art leakage suppression techniques. Traditional leakage error suppression techniques include pulse compensation methods such as DRAG (analytically solved) and GRAPE (numerically solved). Here, we will briefly introduce DRAG as an example.

Considering dominant leakage channels, we chose three energy levels and the Hamiltonian is given by
\begin{equation}
\mathcal{H}_1(t) = \frac{1}{2} \mathbf{B}(t) \cdot \mathbf{S} - \alpha |2\rangle \langle 2|,
\end{equation}
where the operator vector \(\mathbf{S}\) is $S_x = \sum_{k=0,1} \sqrt{k\!+\!1} (|k+1\rangle\langle k| + |k\rangle\langle k+1|)$, $S_y = \sum_{k=0,1} \sqrt{k+1} (i|k+1\rangle\langle k| - i|k\rangle\langle k+1|)$, and $S_z = \sum_{k=0,1,2} (1 - 2k) |k\rangle\langle k|$. \(\alpha\) is the intrinsic anharmonicity of the target transmon qubit. \(\mathbf{B}(t) = \mathbf{B}_0(t) + \mathbf{B}_d(t)\) is the vector of the total microwave field, containing the original microwave field and the additional DRAG-corrected microwave field term, $\mathbf{B}_0(t) = (B_x, B_y, B_z)
= [\Omega(t) \cos(\phi), \Omega(t)\sin(\phi), -\Delta],
\mathbf{B}_d(t) = (B_{d;x}, B_{d;y}, B_{d;z})
= -\frac{1}{2\alpha} (-\dot{B}_y + B_z B_x, \dot{B}_x + B_z B_y, 0)$.

From the perspective of implementation complexity and computational overhead, our scheme avoids introducing additional suppression pulses without altering the original control framework or the shape of the control pulses. This prevents increases in time and control complexity, thereby significantly mitigating the exacerbation of decoherence effects. This also indicates that, when considering decoherence effects in physical implementation, our scheme itself does not have inherent limitations compared to DRAG and GRAPE. More notably, due to the flexibility of the pulse shape, our scheme demonstrates more natural scalability: it is not only suitable for the precise control of two-level systems but can also be effectively extended to multi-level systems and fixed-coupling based two qubit control scenarios, whereas methods relying solely on pulse shape correction struggle to fully compensate for leakage effects in the latter two cases. Furthermore, in our scheme, the leakage coupling term is associated with the tunable parameters in the target ideal Hamiltonian. The small static offsets (SSO) introduced by the tunable parameters are entirely within experimentally acceptable limits, allowing for natural extension to different quantum platforms. Combined with optimized pulse-shaping control techniques, this approach can synergistically suppress other critical error sources and leakage issues present on various platforms.

Numerical simulations show that, as illustrated in Figs. \ref{F10}(a)(b), for single-qubit NOT gate and Hadamard gate, both DRAG and our scheme can significantly mitigate the impact of leakage errors compared to uncorrected gate fidelity. It is worth noting that our scheme demonstrates performance comparable to that of traditional DRAG to a certain extent, while offering greater cost effectiveness and scalability.

\end{appendices}

\end{document}